\documentclass[aps,prb,superscriptaddress,onecolumn,floatfix,showpacs,amsmath]{revtex4}
\usepackage{graphicx}
\usepackage{latexsym}
\usepackage{psfrag}
\usepackage{dcolumn}
\DeclareFontFamily{U}{cmmz}{}
\DeclareFontShape{U}{cmmz}{m}{n}{<-8><8><9><10>%
  <10.95><12><14.4><17.28><20.74><24.88>cmmz10}{}
\DeclareSymbolFont{mz}{U}{cmmz}{m}{n}
\DeclareMathSymbol\Z{\mathord}{mz}{"5A}
\begin{document}

\title{Nanoscale Smoothing and the Analysis of Interfacial Charge 
       and Dipolar Densities}

\author{ Javier Junquera }
\affiliation{ Departamento de Ciencias de la Tierra y 
              F\'{\i}sica de la Materia Condensada, Universidad de Cantabria, 
              Avda. de los Castros s/n, 39005 Santander, Spain}
\affiliation{ Department of Physics and Astronomy, Rutgers University, 
                  Piscataway, New Jersey, 08854-8019, USA }
\author{ Morrel H. Cohen }
\affiliation{ Department of Physics and Astronomy, Rutgers University, 
                  Piscataway, New Jersey, 08854-8019, USA }
\affiliation{ Department of Chemistry, Princeton University, 
                  Washington Road, Princeton, New Jersey, 08554-1009, USA }
\author{ Karin M. Rabe }
\affiliation{ Department of Physics and Astronomy, Rutgers University, 
                  Piscataway, New Jersey, 08854-8019, USA }
\date{\today}

\begin{abstract}
 The interface properties of interest in multilayers include 
 interfacial charge densities, dipole densities, band offsets, and 
 screening-lengths, among others. Most such properties
 are inaccesible to direct measurements, but are key to understanding
 the physics of the multilayers. 
 They are contained within first-principles electronic structure 
 computations but are buried within the vast amount of
 quantitative information those computations generate. Thus far,
 they have been extracted from the numerical data by heuristic nanosmoothing 
 procedures which do not necessarily provide results 
 independent of the smoothing process.
 In the present paper we develop the theory of nanosmoothing,
 establishing procedures for both unpolarized and polarized systems which 
 yield interfacial charge and dipole densities and band offsets invariant
 to the details of the smoothing procedures when the criteria
 we have established are met. 
 We show also that dipolar charge densities, i. e. the 
 densities of charge transferred across the interface, and screening lengths
 are not invariant. We illustrate our procedure with a toy model in which real,
 transversely averaged charge densities are replaced by sums of Gaussians.
\end{abstract}

\pacs{73.30.+y,73.40.Rw,73.40.Qv,73.90.+f}

\maketitle

\section{Introduction}
\label{sec:intro}

 One of the central problems of physics from the mid nineteenth century
 on has been how to make the transition from a microscopic
 to a macroscopic theory of matter.
 This problem has two aspects.
 The first is the task of deriving the equations governing the macroscopic
 behaviour of matter from the underlying microscopic equations,
 exemplified by the derivation of
 Maxwell's macroscopic equations from the microscopic theory
 of charges and fields in vacuum. 
 This problem is elegantly solved by a coarse-graining procedure which 
 takes an average over ``physically infinitesimal'' regions.
 Such an elementary region is chosen to be 
 small enough to let the average of a quantity follow all the changes
 that are observable at the macroscopic level, but large enough compared 
 with characteristic atomic dimensions for it to 
 contain so many particles that the behaviour of an individual particle
 has a negligible effect on the average quantity. 
 This coarse-graining procedure smooths
 over the atomic-scale fluctuations in physical quantities, leaving
 only the slow spatial variation of their macroscopic components.
 A beautifully clear derivation of the macroscopic Maxwell's equations,
 first derived by H. A. Lorentz in 1902, \cite{Lorentz-02}
 can be found in Rosenfeld's {\it Theory of Electrons},
 a now forgotten classic. \cite{Rosenfeld-51}
 In the coarse-graining procedure there are three distance scales:
 $\lambda_1$, the scale on which the macroscopic quantities vary;
 $\lambda_2$, the scale on which the smoothing is carried out;
 and $\lambda_3$, the microscopic scale, that is, the atomic scale.
 For the procedure to work, $\lambda_1$ must be sufficiently larger
 than $\lambda_3$ that the pair of inequalities
 $\lambda_1 >> \lambda_2 >> \lambda_3$ can both be satisfied. Indeed,
 this criterion makes clear the distinction between
 macroscopic and microscopic.

 The second aspect is the task of deriving macroscopic constitutive 
 equations from microscopic properties. An early example
 is the derivation of the Claussius-Mossotti 
 equation \cite{Clausius-79,Mossotti-50} which provides the 
 link between the microscopic polarizability 
 (response of the atoms or molecules to the local electric field)
 and the macroscopic dielectric constant. 

 The systems of interest in the present paper are those with interfaces
 between quite different materials. 
 A planar capacitor comprised of an insulating layer sandwiched between 
 metallic electrodes is a good example of such a system. 
 Further examples are heterojunctions between different semiconductors 
 and Schottky barriers at semiconductor-metal interfaces.
 In all such systems, there are multiple causes of charge transfer
 across or to the interfaces. These can include charge transfer
 to establish spatial uniformity of the chemical potential (the Fermi level),
 charge transfer in response to the local change in chemical 
 composition across the interface and to the interface-induced atomic
 relaxation of the structure,
 charge accumulation to screen the interface charge density 
 associated with the termination of bulk polarization at the interface, and
 charge accumulation attendant to charging or shorting
 \cite{Ghosez-06,Junquera-03.1} of a capacitor.
 Interface dipole densities arise from such transfer of charges
 and are responsible for offsets of the average electrostatic potentials
 across the interfaces, a dominant factor in determining 
 Schottky barriers and valence and conduction band offsets in
 semiconductor heterojunctions. \cite{Franciosi-96,Peressi-98}
 In all such cases, the charge density of each material is perturbed,
 and the perturbation is localized near the interfaces,
 usually within a few interatomic distances. 

 The case of thin ferroelectric films between metallic electrodes provides
 a good illustrative example of this general class of systems.
 Since the early seventies a phenomenological model 
 has been developed \cite{Batra-72,Wurfel-73,Batra-73,Mehta-73}
 to explain the modification of the polar phases 
 (substantial reduction of the spontaneous
 polarization for small thicknesses, or even the
 complete suppression of ferroelectricity below a certain
 critical thickness) and of their thermodynamic properties (depression of
 the transition temperature with respect to that of the
 bulk material).
 The model, mainly due to Batra and coworkers, 
 relies on three basic assumptions:
 $(i)$ the polarization charge lies in a sheet
 right at the interface,
 $(ii)$ the surface polarization charge density equals the
 magnitude of the polarization inside the thin film,
 and ($iii$) the free compensation charge
 spreads out at least over a finite distance
 $\lambda$ within the electrode, decaying exponentially
 towards its interior as in the Thomas-Fermi approximation.
 In this model, the screening length $\lambda$ is dependent only on
 intrinsic properties of the electrode, such as the density
 of free carriers or the dielectric constant.
 All effects are neglected which might come from a particular choice of the 
 electrode/ferroelectric interface, such as the different chemical bondings
 formed at the junction or the interpenetration
 of the electrode and dielectric/ferroelectric wave functions
 that might screen the polarization charge in part within the
 insulator, reducing therefore the magnitude of the interface dipole density.
 Atomic level charge fluctuations are neglected as well, 
 implying a smoothing on a microscopic scale distinct from the smoothing on the
 coarse-graining scale in the derivation of macroscopic equations and
 properties.

 The need to go beyond such simple models is well illustrated by
 this interpenetration of wave functions across the interface.
 As pointed out first by Heine, \cite{Heine-65} and 
 later by Tejedor {\it et al.} \cite{Tejedor-77} and 
 Tersoff, \cite{Tersoff-84.2} the bulk Bloch-states of the metal with
 energies below the Fermi level of the metal and within the
 semiconductor band-gap and its valence band decay exponentially 
 inside the semiconductor
 (and, indeed, might have a significant amplitude for a few layers from 
 the interface), creating a continuum of gap states [the so-called
 metal-induced gap states (MIGS)]. 
 Achieving deep understanding of interface properties with 
 quantitative predictive power and free of adjustable parameters 
 when they are determined by such effects occurring at the atomic scale
 requires first-principles computations.
 In recent years it has become possible to carry out first-principles
 calculations for systems of the complexity of those 
 under discussion here.
 These simulations provide a wealth of
 information at the atomic level 
 about the structural and electronic
 properties of materials and their responses to various external 
 perturbations. \cite{Martin} 
 Some quantities, such as the microscopic charge density distribution
 $\rho\left( \vec{r} \right)$ or the corresponding electrostatic
 potentials, are routinely available from first-principles calculations.

\begin{figure}[htbp]
    \begin{center}
  \includegraphics[width=10cm] {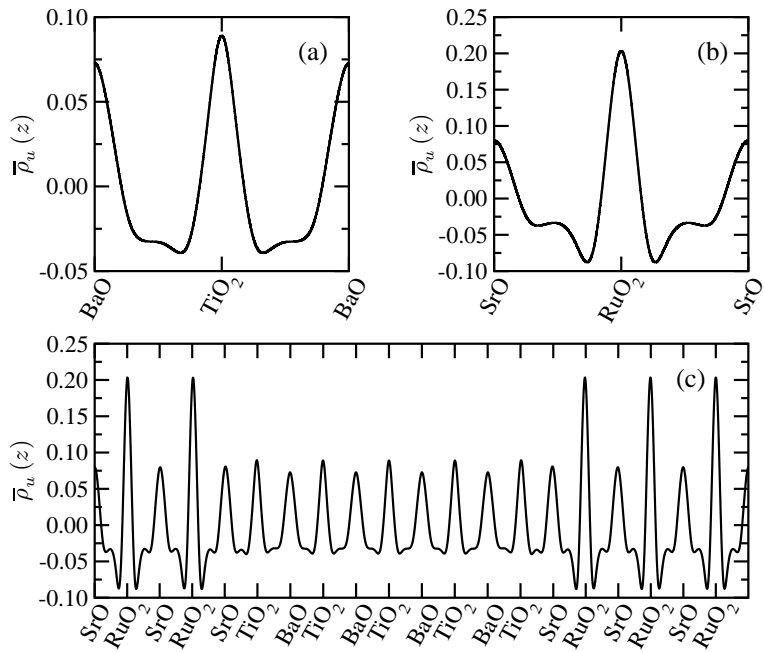}
       \caption{ Charge density laterally averaged parallel to
       $\left( 001 \right)$ planes
       [see Eq. (\ref{eq:planarav}) below] of bulk 
       BaTiO$_3$ (a), SrRuO$_3$ (b), and (c) an unpolarized 
       planar capacitor comprised
       of 4 layers of BaTiO$_3$ between metallic electrodes 
       made of 5 layers of SrRuO$_3$, all from first-principles.  
       The charge-density profile of the capacitor 
       looks like a juxtaposition of the bulk charge densities of
       the two materials, highlighting the fact that 
       the charge transferred from one material to the other
       to establish the constancy of the chemical potential is
       overwhelmed by much larger variations of the total microscopic charge
       density. Details of the first-principles calculations
       can be found in Refs. \onlinecite{Junquera-03.1} 
       and \onlinecite{Ghosez-06}.
       The unit of the charge density is electrons/bohr$^3$. 
         }
       \label{fig:chargedensity}
    \end{center}
 \end{figure}

 The question becomes how to extract from the immense detail
 provided by the first-principles computations reliable values of the physical
 quantities of interest {\bf --} interface charge and dipole
 densities, screening
 lengths, etc. {\bf --} which enter the pseudo-macroscopic
 models currently used.
 Two major difficulties arise.
 First, coarse graining is inapplicable because the relevant distance 
 scale for interface properties is the atomic scale, i. e. 
 $\lambda_{2} \sim \lambda_{3}$. 
 Second, the charge-density changes associated with maintaining
 the constancy of the Fermi levels can be orders of magnitude smaller
 than the unperturbed bulk charge densities, themselves very
 rapidly varying functions of position, reflecting the underlying 
 atomic structure. 
 The relative magnitudes of these changes in the microscopic charge densities 
 are illustrated in Fig. \ref{fig:chargedensity}.
 Moreover, the polarization-induced charge densities and their screening
 charge densities can be smaller by additional orders of magnitude
 than those arising from imposing Fermi-level constancy. 
 Therefore, all the interface-related dipole densities are
 overwhelmed by much larger variations of the total microscopic charge
 density. 

 \psfrag{Yaxis1}[cc][cc]{$\overline{\rho}_{u} \left( z \right)$}
 \psfrag{Yaxis2}[cc][cc]{$\Delta \overline{\rho}_{u} \left( z \right)$}
 \psfrag{Yaxis3}[cc][cc]{$\overline{\overline{\rho}}_{u} \left( z \right)$}
 \psfrag{Yaxis4}[cc][cc]{$g^{(s)}_{i} \left( z - z^{s}_{i} \right)$}

 It is still possible to carry out smoothing of the computed
 charge density at the nanoscale where the conditions for
 coarse-graining are not met, providing sufficient care is taken.
 A heuristic smoothing procedure has been 
 introduced \cite{Baldereschi-88,Colombo-91,Peressi-98} to
 extract the quantities of interest from the results of first-principles
 charge-density computations as described in more detail below.
 However, that procedure has not yet been systematically 
 analyzed to establish the
 conditions under which it accurately extracts the quantities 
 of interest:
 surface charge densities, surface dipole
 densities, etc. 
 In the present paper, we introduce such an analysis.
 In addition, in order to focus only on the perturbations introduced
 by the nanosmoothing procedure, avoiding other sources of numerical noise
 coming from the first-principles simulations, we illustrate the
 analysis with a toy model whose accuracy can be arbitrarily improved.
 The application of this theory to first-principles computations on
 realistic ferroelectric capacitors is the subject of a forthcoming paper. 
 Despite this focus on the ferroelectric capacitor, our analysis is 
 of general utility for the
 extraction of interface properties for all multilayer systems. 

 The rest of the paper is organized as follows. 
 We set the grounds of our discussion of 
 simulations of interfaces from first-principles  
 and define the 
 microscopic behaviour of the different quantities that are the
 targets of our study in
 Sec. \ref{sec:first-principles}. 
 In Sec. \ref{sec:unpolar} we define the interface quantities
 of interest for an unpolarized interface. 
 In Sec. \ref{sec:refint} we describe the difficulties encountered in
 defining precisely the location of a reference interface.
 We develope the theory of nanosmoothing of unpolarized systems
 in Sec. \ref{sec:nanosmoothing} with particular attention to questions 
 of the sensitivity of quantities of interest to the smoothing procedure.
 We describe in Sec. \ref{sec:toymodel} a toy model used
 to illustrate all of the essential elements of the smoothing theory.
 In Sec. \ref{sec:resultsunpolar}, we present 
 the results of the toy model computations
 for the unpolarized case. We generalize the nanosmoothing theory of
 Sec. \ref{sec:nanosmoothing} for polarized systems in Sec. \ref{sec:polar} 
 and present the corresponding toy model results in Sec. \ref{sec:resultspolar}.
 Finally in Sec. \ref{sec:summary}, 
 we summarize our results, emphasizing the specific
 criteria smoothing functions must satisfy to yield interfacial properties
 insensitive to their parameters and specifying which of our results are new.

\section{Simulation of interfaces from first principles.}
\label{sec:first-principles}

\begin{figure}[htbp]
    \begin{center}
    \includegraphics[width=6cm] {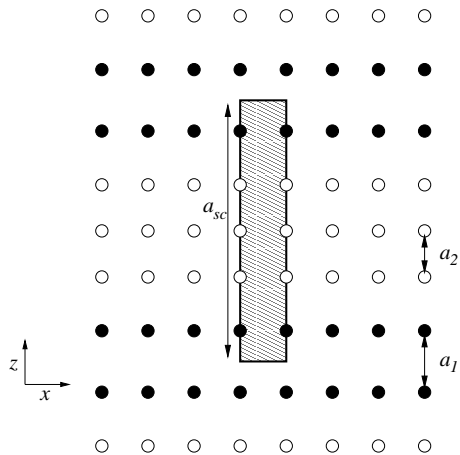}
       \caption{ Schematic view of a supercell used in 
                 first-principles simulations of interfaces.
                 The $z$ axis is taken as normal to the interface, whereas
                 the plane parallel to the discontinuity is taken as the 
                 $(x,y)$ plane. 
                 $n_{1}$ and $n_{2}$ multiatomic layers of two materials,
                 whose bulk lattice constants in the
                 out-of-plane direction are respectively $a_{1}$ and $a_{2}$,
                 are stacked to build a basic unit cell (hatched rectangle)
                 that is periodically repeated in space.
                 $a_{sc}$ is the length of the supercell in the 
                 out-of-plane direction, with
                 $a_{sc} = n_{1} a_{1} + n_{2} a_{2}$.
                 In the figure, $n_{1}$ = 2, and $n_{2}$ = 3.
         }
       \label{fig:supercell}
    \end{center}
 \end{figure}

 First-principles calculations of interfaces between two materials,
 where there is no periodicity in at least one direction,
 are almost universally done by means of the supercell 
 approximation. \cite{Payne-92}
 Within this approach a basic unit cell that contains a suitable
 number of multiatom layers of the
 two materials is periodically repeated over all space
 (Fig. \ref{fig:supercell}).
 For the interfaces within a nanostructured multilayer to be well defined,
 with properties distinct from those of the bulk-like regions between them,
 the widths of the layers of each material introduced in the
 construction of the basic unit cell must be large enough to avoid the
 interaction between adjacent interfaces through the bulk materials, so that
 the calculation accurately represents an isolated interface.

 Throughout this work, we shall assume that the interface is oriented
 along the $z$ axis, and each material is periodic in the plane
 parallel to the interface, referred to as the $(x,y)$ plane.

 Precisely the same methodology is used to treat
 nanostructured multilayer materials, and in this paper
 we do not distinguish between the cases, specializing to multilayers
 in which the individual material thicknesses are large enough 
 for the interfaces to be noninteracting.

 The microscopic charge densities $\rho \left( \vec{r} \right) $
 [see Fig. \ref{fig:chargedensity}(c)]
 and electrostatic potentials $V \left( \vec{r} \right) $ provided by the
 first-principles computations for the previously described supercells
 are continuous functions periodically repeated in space
 with the periodicity of the supercell ($a_{sc}$ in Fig. \ref{fig:supercell}).
 A few interatomic distances away from the interfaces, the microscopic
 quantities recover their bulk features.
 In other words, if the
 layer widths are large enough, it is possible
 to identify ``bulk-like'' regions in the middle of each of the layers that 
 constitute the superlattice, with large variations of the microscopic
 quantities with the same periodicity as in the bulk, reflecting the
 underlying atomic structure.

 Many of the interface-related quantities that we shall define below
 depend only on what happens in the direction perpendicular to the
 interface, where the actual discontinuities of the physical structure occur. 
 Since we assume in-plane periodicity, we can trivially eliminate the
 in-plane dependence by taking a planar average of the corresponding
 microscopic quantity, e. g., for the electron density,

 \begin{equation}
    \overline{\rho} \left( z \right) =
    \frac{1}{S} \int \int_{S} \rho \left( \vec{r} \right) \: dx \: dy, 
    \label{eq:planarav}
 \end{equation}
 
 \noindent where $S$ is the area of the interface unit cell. 
 
 As is proven in Appendix \ref{sec:transversepoisson}, 
 this transverse averaging 
 has no effect on the Poisson equation,
 which reads after averaging

 \begin{equation}
    \overline{\nabla^{2} V \left( \vec{r} \right) }  =
    \nabla^{2} \overline{V} \left( z \right) = 
    \frac{d^{2} \overline{V} \left( z \right) }{dz^{2}} =
    - 4 \pi \overline{ \rho } \left( z \right). 
    \label{eq:transavpoisson}
 \end{equation}

 \section{Interface quantities of interest; The unpolarized case.}
 \label{sec:unpolar}

 One of the most important physical properties of heterojunction
 devices is the band offset or Schottky barrier at an interface, that is,
 the relative positions of the energy levels on both sides of the interface.
 In the case of semiconductor heterojunctions, the
 valence-band offset (VBO) [conduction-band offset (CBO)]
 is defined as the difference between the positions of the
 tops of the valence bands [the bottoms of the conduction bands] of the
 two materials.
 In the case of metal-semiconductor contacts, we can define the $p$-type 
 ($n$-type) Schottky barrier as the difference between the Fermi level
 of the metal and the top of the valence band (bottom of the conduction
 band) of the semiconductor. 
 These differences in the band positions determine the effective barrier
 for electron or hole transport across the junction.

 The computation of such effects from first-principles
 cannot be achieved by a direct comparison of the corresponding
 single-particle energies (tops of the valence bands, bottoms of the
 conduction bands and/or Fermi level of the metal) 
 in the two compounds as obtained from two independent
 {\it bulk} band-structure calculations.
 The reason is the lack of an intrinsic energy origin to which
 to refer all the energies: in a first-principles simulation,
 the Hamiltonian eigenvalues are referred to an average of the electrostatic
 potential that is ill defined for an infinite system \cite{Kleinman-81}
 where, 
 due to the long-range nature of the Coulomb interaction,
 it is defined only to within an arbitrary constant.
 Consequently, together with the eigenvalue difference, we must consider
 both the shift of this average between the two materials and 
 the redefinition of the averaging process so that it is 
 appropiate to the multilayer system under consideration.
 As was mentioned in Sec. \ref{sec:intro}, 
 the coarse-graining procedure used historically fails when applied
 to the results of first-principles computations. 
 We designate the averaged
 potential as $\langle V_{u} \rangle$ and its shift as
 $\Delta \langle V_{u} \rangle$. The brackets $\langle \rangle$ 
 indicate that the averaging process is not yet defined.
 We shall label all quantities of 
 physical interest of the unpolarized system by a subscript $u$.
 This potential shift depends on the dipole induced 
 by the electronic charge transferred
 from one side of the interface to the other after interfacial
 hybridization. 
 As the charge transfer depends not only on the materials that
 constitute the interface, but also on intrinsic interface effects
 such as the chemical composition (termination of each material
 at the interface), on particular orientation and on other 
 structural details, the shift can only be obtained from 
 a self-consistent calculation on a supercell including both materials.
 This ensures that the averaged electrostatic potentials of both materials
 in the ``bulk-like'' regions defined in the previous section,
 where all the physical quantities recover the bulk features,
 are expressed with respect the same reference and allows a direct
 extraction of the shift.

 This shift of the averaged electrostatic potential should be directly related
 to an averaged interface dipole-moment density $\langle p_{u} \rangle$,

 \begin{equation}
    \Delta \langle V_{u} \rangle = 4 \pi \langle p_{u} \rangle. 
    \label{eq:deltavdipole}
 \end{equation}

 \noindent We prove in Appendix \ref{sec:elecdipole} 
 that Eq. (\ref{eq:deltavdipole})
 holds for the specific definition of the average procedure 
 $\langle \rangle$ introduced in Sec. \ref{sec:nanosmoothing-procedure} 
 below and discussed in this context in Sec. \ref{sec:nanosmoothing-poisson}.

 From a fundamental point of view, \cite{Jackson} 
 the charge $Q_{u}$, and electric dipole moment at the
 interface $p_{u}$ are defined respectively as the zero and the first moment 
 of the total microscopic charge density, 

 \begin{subequations}
    \begin{align}
      Q_u & = \int_{z_{1}}^{z_{2}} dz \:\: \overline{\rho}_u \left( z \right), 
      \label{eq:qu} 
      \\
      p_u & = \int_{z_{1}}^{z_{2}} dz \:\: z \overline{\rho}_u \left( z \right).
      \label{eq:pu}
    \end{align}
 \end{subequations}

 \noindent For the previous equations to be meaningful and truly 
 represent an interface quantity, the thicknesses
 of the adjacent layers must be
 wide enough so that they contain regions within which the charge density
 is essentially unaffected by the presence of interfaces 
 within which $z_{1}$ and $z_{2}$ could be located.

 However, from a practical point of view, such a definition poses
 serious questions.
 Indeed, both $Q_{u}$ and $p_{u}$ are ill-defined
 due to the large and rapid oscillations of 
 the microscopic charge density [see Fig. \ref{fig:chargedensity}(c)].
 In fact, different choices of the 
 integration limits $z_{1}$ and $z_{2}$ yield widely different values of 
 the charge and interface dipole moment.
 Only in the extreme case of a Clausius-Mossotti model,
 in which the total charge is unambiguously decomposed into an assembly of
 localized and neutral charge distributions, so that a unit cell 
 can be chosen with no charge at the surface, the dipole moment of a periodic 
 charge distribution would be well-defined
 as the integral of the first moment of the charge density. 
 However, any Claussius-Mossotti
 approach does not correspond to reality, particularly in materials
 where delocalized covalent charge is present.\cite{Resta-94}
 This ambiguity in the definition of $p_{u}$ with respect to
 the boundaries of the region within which the dipole moment 
 is computed
 is closely connected to the problem of defining the polarization
 of a periodic system from the charge density. \cite{Martin-74}

 \section{The difficulty of defining a reference interface.}
 \label{sec:refint}

 In order to get rid of bulk effects and extract interface-related features,
 some authors \cite{Bylander-87.1,Bylander-87.2,Bylander-88.1,Bylander-88.2} 
 have defined an ideal interface by stacking alternate slabs,
 each of them made from slicing the planar average of the
 bulk charge density of the
 corresponding material perpendicular to a particular direction.
 Let us define $\overline{\rho}_{0}^{(1)} \left( z \right)$
 and $\overline{\rho}_{0}^{(2)} \left( z \right)$
 as the planar-averaged [Eq. (\ref{eq:planarav})]
 bulk charge densities of the left and right material 
 respectively, unaffected by the presence of the interface.
 They are locally periodic in $z$, and, assuming that both bulk
 materials are unpolarized, the net charge and
 the dipole moment vanish within each bulk unit cell,

 \begin{subequations}
    \begin{align}
       & \int_{a_{bulk}^{(s)}} dz \:\:
         \overline{\rho}_{0}^{(s)} \left( z \right) = 0,
       \label{eq:chargerho0}     
       \\
       & \int_{a_{bulk}^{(s)}} dz \:\: 
         z \overline{\rho}_{0}^{(s)} \left( z \right) = 0,
       \label{eq:dipolerho0}     
    \end{align}
 \end{subequations}

 \noindent where $a_{bulk}^{(s)}$ is the lattice constant 
 of the bulk unit cell 
 of each material in the $z$ direction, 
 and $s$ refers to the side of the interface 
 considered, 1 or 2. The bulk unit cell boundaries are chosen 
 so as to preserve inversion symmetry.

 Replicating the bulk charge densities up to the as yet
 unspecified interface from each side,
 we could define the reference charge density 
 $\overline{\rho}_{0}$ for all $z$ as

 \begin{eqnarray}
     \overline{\rho}_{0} \left( z \right)  =  
     \left\{ \begin{array}{cl}
        \overline{\rho}_{0}^{(1)} \left( z \right), &  \: \:  z < z_{int},  \\
                                                         \\
        \frac{1}{2}\left[ \overline{\rho}_{0}^{(1)} \left( z_{int} \right)+ 
                        \overline{\rho}_{0}^{(2)} \left( z_{int} \right) \right]
                          , &  \: \:  z = z_{int},                        \\
                                                          \\
        \overline{\rho}_{0}^{(2)} \left( z \right), &  \: \:  z > z_{int},  \\
                                                          \\
        \end{array} \right.
    \label{eq:rho0}
 \end{eqnarray}

 \noindent where $z_{int}$ is the coordinate assigned to the position of
 the interface. $\overline{\rho}_{0} \left( z \right)$ is discontinous
 at the interfacial plane $z_{int}$ by its very definition.

 We now define the interface-induced deformation of the charge
 density $\Delta \overline{\rho}_{u} \left( z \right)$ as

 \begin{equation}
    \Delta \overline{\rho}_{u} \left( z \right) = 
    \overline{\rho}_{u} \left( z \right) - 
    \overline{\rho}_{0} \left( z \right).
    \label{eq:deltarho}
 \end{equation}

 \noindent If the thicknesses of the two layers are wide enough,
 $\Delta \overline{\rho}_{u} \left( z \right)$ becomes negligibly 
 small over the ranges $R_{1}$ in the left material and $R_{2}$ in
 the right material [see Fig. \ref{fig:nanotoy}(b)]. 
 The interface region can thus be identified
 as comprised of those ranges where 
 $\Delta \overline{\rho}_{u} \left( z \right)$
 differs significantly from zero, in other words
 where the microscopic charge density differs from the relevant
 bulk values.
 The interface charge and dipole density associated with
 $\Delta \overline{\rho}_{u} \left( z \right)$ are defined as, 

 \begin{subequations}
    \begin{align}
      \Delta Q_{u} & = \int_{z_{1}}^{z_{2}} dz \:\: 
                  \Delta \overline{\rho}_{u} \left( z \right),
      \label{eq:deltaq}
      \\
      \Delta p_{u} & = \int_{z_{1}}^{z_{2}} dz \:\: 
                 z \Delta \overline{\rho}_{u} \left( z \right).
      \label{eq:deltap}
    \end{align}
 \end{subequations}

 \noindent The advantage of this approach is that both 
 $\Delta Q_{u}$ and $\Delta p_{u}$ are well defined quantities
 with respect the location of the integration limits $z_{1}$ and 
 $z_{2}$ in Eqs. (\ref{eq:deltaq})-(\ref{eq:deltap}),
 provided that $z_{1}$ lies in $R_{1}$ and
 $z_{2}$ lies in $R_{2}$.

 However, this approach has pitfalls.
 In particular $(i)$ the position of the interface, $z_{int}$ in 
 Eq. (\ref{eq:rho0}), is not yet specified. 
 Some recipes have been given for how to cut the bulk slabs,
 but they have limited applicability. One case is for 
 common anion heterostructures
 with non-relaxed interfaces along high-symmetry planes, such as 
 the (001) \cite{Bylander-87.1}, (110) \cite{Bylander-87.2}, 
 or (111) \cite{Bylander-88.1} interfaces of GaAs/AlAs superlattices.
 As soon as an interface-induced rippling of the atomic layers
 is introduced, for instance after an atomic relaxation of the
 interface geometry, the problem of defining the position of the
 interface worsens.
 $(ii)$ Therefore, the interface charge and dipole densities are not unique, 
 since they depend
 critically on where the mathematical surface representing the
 interface is chosen, and no objective criterion for locating it 
 has been established.
 In particular $\Delta Q_{u}$ and $\Delta p_{u}$ 
 [Eqs. (\ref{eq:deltaq})-(\ref{eq:deltap})]
 equal $Q_{u}$ and $p_{u}$ 
 [Eqs. (\ref{eq:qu})-(\ref{eq:pu})]
 if and only if $z_{int} - z_{1}$ contains an integer number
 of unit cells of the left material and $z_{2}-z_{int}$ contains
 an integer number of unit cells of the right material.
 The location of the interface determines where on one side, the bulk charge
 density of the left material is subtracted, and on the
 other side that of the right material.
 A different choice of the mathematical interface can produce 
 very different charge and dipole densities. 
 In addition, comparison between different
 interface orientations makes little sense with this definition.
 \cite{Franciosi-96}
 $(iii)$ Since the interface dipole moment is dependent on
 the reference charge density, the corresponding potential drop
 at the interface ($\Delta_{dipole}$ in the notation of 
 Refs. \onlinecite{Bylander-87.1,Bylander-87.2, Bylander-88.1, Bylander-88.2})
 must be too. However, it is important to note that the 
 potential drop
 generated by $\Delta \overline{\rho}_{u} \left( z \right)$
 is only part of the total potential shift.
 The total charge density of the interface is given by

 \begin{equation}
    \overline{ \rho}_{u}  \left( z \right) = 
    \overline{ \rho}_{0}  \left( z \right) +
    \Delta \overline{ \rho}_{u} \left( z \right),  
 \end{equation}

 \noindent thus the potential shift associated with
 $\overline{\rho}_{0} \left( z \right)$ must be included as well.
 Although the existence of a potential shift generated by 
 $\overline{\rho}_{0} \left( z \right)$ is general,
 we shall explain its origin only in the particular case
 where $z_{int}-z_{1}$ contains an integer number of unit cells 
 of the left material,
 and $z_{2}-z_{int}$ contains an integer number of unit cells of 
 the right material.
 In this particular situation both slabs used to 
 construct the reference charge density in Eq. (\ref{eq:rho0})
 have neither a charge nor a dipole moment.
 Under these circumstances, the potential shift is 
 the difference in the locally averaged potentials
 produced by the $\overline{\rho}_{0}^{(s)}$ of each material in 
 the regions $R_{1}$ and $R_{2}$ respectively.
 For this shift to be an interface property, the layer width of
 each material must be large enough that the local average 
 $\langle V^{(s)}_{slab} \rangle$ has approached the unperturbed 
 value within the
 center of an slab.
 The planar averaged potential within any point of the slab
 will be given by \cite{Kleinman-81} 

 \begin{equation}
    V_{0}^{(s)} \left( z \right) = 2 \pi
      \int dz^{'} | z - z^{'} | \overline{\rho}_{0}^{(s)} (z^{'}),
 \end{equation}

 \noindent so the local average can be computed as

 \begin{equation}
    \langle V^{(s)}_{slab} \rangle = \frac{2 \pi}{a^{(s)}_{bulk}} 
                 \int_{central\:\:cell} d z
                 \int dz^{'} 
                 | z - z^{'} | \overline{\rho}_{0}^{(s)} (z^{'}).
 \end{equation}

 \noindent Since $\langle V^{(s)}_{slab} \rangle$ depends on 
 the charge density distribution,
 it differs for the left and right slabs used in 
 the construction of the reference charge density and produces an
 additional shift.
 Therefore, the total potential drop at the interface is the
 sum of the potential drop generated by 
 $\Delta \overline{\rho}_{u} \left( z \right)$, $\Delta_{dipole}$,
 and the difference of the average potential of the two reference
 slabs $\Delta_{ref}$.
 Only the sum $\Delta_{ref} + \Delta_{dipole}$ is independent
 of the reference charge density chosen and is a physically measurable 
 property of the interface.
 Since each term in the sum is sensitive to the arbitrary location
 of the interface, each must be computed accurately enough for the 
 sensitivity to disappear from the sum. 
 As the charge density shifts of interest are so small, this is an
 unnecessary burden, removed by the use of a proper nanosmoothing
 \cite{Baldereschi-88} procedure as shown in Sec. \ref{sec:nanosmoothing}.

 \section{Nanosmoothing.}
 \label{sec:nanosmoothing}

 \subsection{The procedure.}
 \label{sec:nanosmoothing-procedure}
 
 A procedure to eliminate charge fluctuations 
 in the regions of the material which
 do not contribute to the interfacial hybridization,
 thereby localizing the physically relevant charge densities to the interface,
 consists of filtering out the periodic
 oscillations of microscopic quantities, which typically follow the
 underlying atomic structure, preserving only those features
 that emerge in the vicinity of the interface. 

 To obtain this smoothed charge density,
 we have followed the recipe given by Baldereschi
 {\it et al.} in Ref. \onlinecite{Baldereschi-88}
 and generalized by Colombo and coworkers for lattice mismatched
 heterostructures in Ref. \onlinecite{Colombo-91}.
 Starting from the planar-averaged charge density
 $\overline{\rho}_{u}\left( z \right)$,
 we construct the smoothed density 
 $\overline{\overline{\rho}}_{u}\left( z \right)$
 by convoluting it with a smoothing function $f \left( z \right)$,

 \begin{equation}
    \overline{\overline{\rho}}_{u}\left( z \right) =
              \int dz^{'} f ( z - z^{'} )
              \overline{\rho}_{u} ( z^{'} )
    \label{eq:nanosmoothing}
 \end{equation}

 \noindent which has the following properties: \cite{Ashcroft}

 \begin{subequations}
    \begin{align}
       & f \left( z \right)           >   0, \hspace{1.5cm} \left|z \right| < L,
       \label{eq:filterfuna}
       \\
       & f \left( z \right)           =   0, \hspace{1.5cm} \left|z \right| \ge L,
       \label{eq:filterfunb}
       \\
       & f \left( -z \right)          =   f   \left( z \right) , 
       \label{eq:filterfunc}
       \\
       & \int dz f \left( z \right)   = 1,     
       \label{eq:filterfund}
    \end{align}
 \end{subequations}

 \noindent so that

 \begin{equation}
    \overline{\overline{\rho}}_{u}\left( z \right) =
    \int_{z-L}^{z+L} d z^{'} f ( z - z^{'} )
    \overline{\rho}_{u} ( z^{'} ).
    \label{eq:bbrho}
 \end{equation}

 \noindent In addition, $f \left( z \right)$ should be monotonic in
 $\left| z \right|$ and sufficiently smooth itself.
 $L$ should be chosen on the scale of the unit cell length or larger,
 but smaller than the widths of the left and right layers. A sharper
 criterion for $L$ is introduced below.
 We define this as the averaging procedure $\langle \rangle$ left
 unspecified above in Sec. \ref{sec:unpolar}.

 The particular smoothing function we have used, following
 Refs. \onlinecite{Peressi-98}, 
 \onlinecite{Baldereschi-88} and \onlinecite{Colombo-91},
 is the convolution of two square-wave filter functions:

 \begin{equation}
   f ( z - z^{'} ) =
   \int dz^{''} \omega_{l_{1}} ( z-z^{''} )
   \omega_{l_{2}} ( z^{''}-z^{'} ) ,
   \label{eq:filterconvolution}
 \end{equation}

 \noindent where

 \begin{subequations}
    \begin{eqnarray}
      \omega_{l} \left( z \right)
      & = & \frac{1}{l} \Theta\left( \frac{l}{2} - |z| \right),
      \label{eq:step1}
      \\
      \Theta\left( z \right) & = & \left\{ \begin{array}{c}
                                           1, \: \: \: \: z > 0   \\
                                                                    \\
                                           0, \: \: \: \: z \le 0 .   \\
                                           \end{array} \right.
    \label{eq:step2}
    \end{eqnarray}
 \end{subequations}

 \noindent Giustino and coworkers \cite{Giustino-03,Giustino-05} propose
 convolution with a Gaussian kernel that can be an approximation to the
 asymptotic limit of a convolution of a large number of square-wave
 filter functions. This method is best suited for superlattices where
 crystal deviates from perfect periodicity far away from the interface
 so that it is not possible to define regions where the interface-induced 
 charge density vanishes, 
 or in disordered three-dimensional systems with 
 short-range order. 
 Even more general functions can be used, providing the criteria
 established above are met.

 The explicit dependence of $f \left( z \right)$, defined in 
 Eq. (\ref{eq:filterconvolution}), on $z$ is:

 \begin{eqnarray}
     f \left( z \right)  =  \left\{ \begin{array}{cl}
                  0, &  \: \: \left| z \right| > \frac{l_{1} + l_{2}}{2},  \\
                                                          \\
                  \frac{1}{l_{1} l_{2}} \left[ \frac{l_{1} + l_{2}}{2} -
                          \left| z \right| \right],
                  &  \: \: \frac{\left| l_{1} - l_{2} \right|}{2} <
                        \left| z \right| \le
                        \frac{ l_{1} + l_{2} }{2},
                                                           \\
                  \frac{1}{l_{>}} ,
                  &  \: \: 0 < \left| z \right| <
                  \frac{\left| l_{1} - l_{2} \right|}{2} ,
                                     \end{array} \right.
   \label{eq:doublefilter}
 \end{eqnarray}

 \psfrag{Le1}[cc][cc]{$\frac{1}{l_{2}}$}
 \psfrag{Le9}[cc][cc]{$\frac{1}{l_{1}}$}
 \psfrag{Le8}[cc][cc]{$\frac{1}{l_{>}}$}
 \psfrag{Legend1}[cc][cc]{$-\frac{l_{1}+l_{2}}{2}$}
 \psfrag{Legend4}[cc][cc]{$\frac{l_{1}+l_{2}}{2}$}
 \psfrag{Legend2}[cc][cc]{$-\frac{\left|l_{1}-l_{2}\right|}{2}$}
 \psfrag{Legend3}[cc][cc]{$\frac{\left|l_{1}-l_{2}\right|}{2}$}
 \psfrag{Legend5}[cc][cc]{$-\frac{l_{1}}{2}$}
 \psfrag{Legend8}[cc][cc]{$\frac{l_{1}}{2}$}
 \psfrag{Legend6}[cc][cc]{$-\frac{l_{2}}{2}$}
 \psfrag{Legend7}[cc][cc]{$\frac{l_{2}}{2}$}

 \begin{figure}[htbp]
    \begin{center}
       \includegraphics[width=10cm,angle=0] {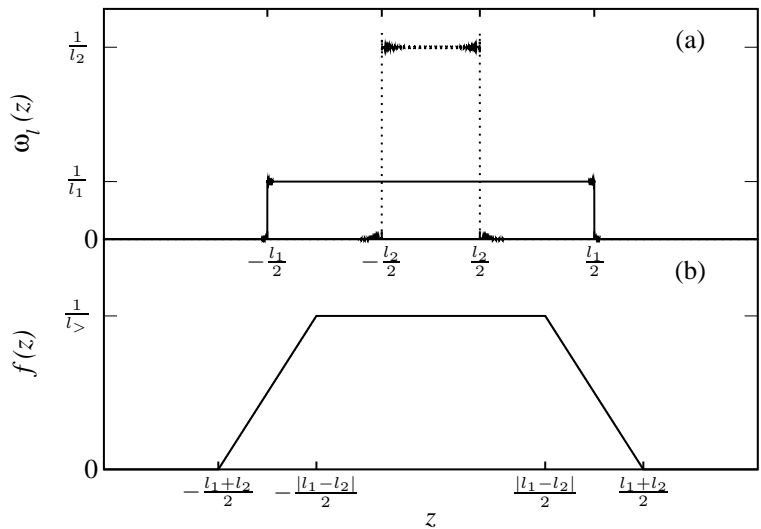}
       \caption{Filter functions used for the smoothing of the charge density
                and potential in the present work.
                Two square wave functions of lengths
                $l_{1}$ and $l_{2}$,
                as defined in
                Eqs. (\ref{eq:step1}) and (\ref{eq:step2}),
                are represented in panel (a).
                The corresponding convolution, given by
                Eq. (\ref{eq:filterconvolution}),
                is shown in panel (b). For numerical calculations, the
                square-wave functions are expanded in Fourier series.
                The small oscillations in the vicinity of each discontinuity
                are due to the Gibbs phenomenon. \cite{Arfken}
               }
       \label{fig:filter}
    \end{center}
 \end{figure}

 \noindent as shown in Fig. \ref{fig:filter}.
 $l_{>}$ is the greater of $l_{1}$ and $l_{2}$.

 After nanosmoothing the planar averaged charge density
 $\overline{\rho}_{u} \left( z \right)$, the resulting charge
 density $\overline{\overline{\rho}}_{u} \left( z \right)$ 
 is a continuous function that joins smoothly at the interface.
 Even though we have not subtracted any reference charge, 
 the smoothed charge density becomes negligibly 
 small over ranges $R^{'}_{1}$ in the left material and $R^{'}_{2}$ in
 the right material [see Fig. \ref{fig:nanotoy}(c)],
 and the interface region can be 
 unambiguosuly defined as the region
 where the smoothed charge significantly differs from zero. 
 Note that $R_{s}$ and $R^{'}_{s}$  will differ in general,
 as shown in Fig. \ref{fig:diffr} and
 discussed further in Secs. \ref{sec:nanosmoothing-insensitivity}
 and \ref{sec:resultsunpolar}.

 \subsection{The Poisson equation and potential shifts.}
 \label{sec:nanosmoothing-poisson}
 
 Providing the filter function $f(z)$ satisfies the 
 following additional conditions,

 \begin{subequations}
   \begin{align}
     & \frac{d^{2} f(z) }{dz^{2}} \:\: {\rm exists},
     \label{eq:conditionf1}
     \\
     & \left. \frac{d f(z) }{dz} \right|_{z=-L} =  
       \left. \frac{d f(z) }{dz} \right|_{z=+L} = 0,
     \label{eq:conditionf2}
     \\
     & \left. f(z) \right|_{z=-L} =  
       \left. f(z) \right|_{z=+L} = 0.
     \label{eq:conditionf3}
   \end{align}
 \end{subequations}

 \noindent The Poisson equation remains invariant after the nanosmoothing
 (see Appendix \ref{sec:nanosmoothpoisson})
 and transforms into

 \begin{equation}
    \nabla^{2} \overline{\overline{V}}\left( z \right) =
    \frac{d^{2} \overline{\overline{V}} \left( z \right) }{dz^{2}} =
    - 4 \pi \overline{\overline{\rho}} \left( z \right).
    \label{eq:poissonnanosmoothed}
 \end{equation}

 The nanosmoothing function $f (z)$ defined in Eq. (\ref{eq:step1}) and
 Eq. (\ref{eq:step2}) violates condition Eq. (\ref{eq:conditionf1})
 at its end points $z = \pm \frac{l}{2}$, where $f (z)$ is discontinuous
 so that condition (\ref{eq:conditionf3}) cannot be 
 unambiguously applied.
 Similarly, the $f(z)$ of Eq. (\ref{eq:doublefilter}) violates 
 condition (\ref{eq:conditionf1}) at the four points 
 $z = \pm \frac{l_{1} \pm l_{2}}{2}$, and $df/dz$ is discontinuous
 at the points $z = \pm \frac{l_{1} \pm l_{2}}{2}$ so that
 condition (\ref{eq:conditionf2}) cannot be  unambiguously applied.
 Nevertheless, one can think of the $f(z)$ as a {\it distribution}, a 
 family of smooth functions all of which meet conditions 
 (\ref{eq:conditionf1}) - (\ref{eq:conditionf3}) and which approach
 the $f(z)$ of Eq. (\ref{eq:doublefilter}) as their limit.
 In practice, because the smoothing operation is a convolution,
 Eq. (\ref{eq:nanosmoothing}), one carries out smoothing via
 fast Fourier transformations. 
 The family of the finite Fourier series involved is thus 
 a distribution which converges to $f(z)$ in the limit, meeting
 conditions (\ref{eq:conditionf1}) - (\ref{eq:conditionf3}) 
 along the way.

 Eq. (\ref{eq:poissonnanosmoothed}) holds in general for both the unpolarized
 and the polarized cases, so subscripts have been omitted.
 Otherwise the subscript $u$ is used because we are presently treating the 
 unpolarized case. 

 We prove in Appendix \ref{sec:elecdipole} that
 the full electrostatic potential shift $\Delta \overline{\overline{V}}_{u}$
 is given by the nanosmoothed dipole density $\overline{\overline{p}}_{u}$,

 \begin{equation}
    \Delta \overline{\overline{V}}_{u} = 4 \pi \overline{\overline{p}}_{u},
    \label{eq:relshiftdip}
 \end{equation}

 \noindent where 

 \begin{equation}
    \overline{\overline{p}}_{u} = \int_{z_{1}}^{z_{2}} dz \:\:
                 z \overline{\overline{\rho}}_{u} \left( z \right).
    \label{eq:macrodipole}
 \end{equation}

 \subsection{Insensitivity of the dipole-moment density and potential 
             shift to the smoothing function.}
 \label{sec:nanosmoothing-insensitivity}

 The microscopic charge density $\overline{\rho}_{u}$ 
 must return to the bulk microscopic charge densities
 $\overline{\rho}_{0}^{(1)}$ and $\overline{\rho}_{0}^{(2)}$
 in the regions $R_{1}$ and $R_{2}$ for it to be possible to 
 ascribe physical properties specifically to individual
 interfaces. However rapidly it approaches those values,
 the approach must be complete within $R_{1}$ and $R_{2}$,
 as summarized by Eq. (\ref{eq:rhor1r2}),

 \begin{eqnarray}
    \overline{\rho}_{u} \left( z \right)  =  \left\{ \begin{array}{ll}
      \overline{\rho}_{0}^{(1)} \left( z \right),  \: \:  z \in R_{1},  \\
      \\
      g\left( z \right) \ne \overline{\rho}_{0}^{(1)} \left( z \right), 
      \:\: z \not\in R_{1}, \\
      \\
      g\left( z \right) \ne \overline{\rho}_{0}^{(2)} \left( z \right),  
       \: \:  z \not\in R_{2}, \\
      \\
      \overline{\rho}_{0}^{(2)} \left( z \right),  \: \:  z \in R_{2},  \\
                                                           \\
      \end{array} \right.
   \label{eq:rhor1r2}
 \end{eqnarray}

 \noindent with $g\left( z \right)$ such that 
 $\overline{\rho}_{u} \left( z \right)$
 and all its derivatives are continuous (the cusps at the nuclei
 are washed out by lateral averaging).
 Consequently, {\it locally} within $R_{1}$ and $R_{2}$, 
 the microscopic charge density $\overline{\rho}_{u}$ 
 can be {\it represented} by the
 Fourier transforms of $\overline{\rho}_{0}^{(1)}$ 
 and $\overline{\rho}_{0}^{(2)}$,

 \begin{align}
    & \overline{\rho}_{0}^{(s)} \left( z \right) =
    \sum_{n} A^{(s)}_{n} e^{i \kappa^{(s)}_{n} z} ; 
    \nonumber \\
    & \kappa^{(s)}_{n} = \frac{2 \pi n}{a^{(s)}_{bulk}}; \:\: 
    n \in \Z; \:\:
    z \in R_{s}; \:\:
    s = 1, 2.
    \label{eq:ftrho12}
 \end{align}

 \noindent In Eq. (\ref{eq:ftrho12}) $A^{(s)}_{0}$ vanishes because
 of charge neutrality.

 The smoothing function $f ( z - z^{'} )$, Eq. (\ref{eq:filterconvolution}), 
 is a convolution of two square wave functions 
 $\omega_{l_{s}} ( z - z^{'} )$, 
 Eqs. (\ref{eq:step1})-(\ref{eq:step2}), s = 1, 2.
 The order of the $\omega_{l_{s}}$ within the 
 convolution is immaterial,
 so the $\omega_{l_{s}}$ can be applied to the nanosmoothing of
 $\overline{\rho}_{0}^{(s)}$ first, in the two-step nanosmoothing process
 implied by use of $f ( z - z^{'} )$ in Eq. (\ref{eq:nanosmoothing}).
 As long as $l_{s}$ is an integer multiple of the lattice constant
 of material $s$, all contributions to $\overline{\rho}_{0}^{(s)}$ from
 $A^{(s)}_{n}, n \ne 0$, are smoothed to zero, leaving only 
 $A^{(s)}_{0}$ which itself vanishes.
 However the regions $R^{'}_{s}$ within which $\overline{\overline{\rho}}_{u}$
 vanishes lie within $R_{s}$ because smoothing $\overline{\rho}_{u}$ 
 within $R_{s}$ brings into $\overline{\overline{\rho}}_{u}$
 values of $\overline{\rho}_{u}$ for $z$ outside $R_{s}$.
 The limits $z_{s}$ for the determination of $\overline{\overline{p}}$
 in Eq. (\ref{eq:macrodipole})
 must lie within $R^{'}_{s}$, and $L$ must be significantly smaller than 
 the width of $R_{s}$.

 Similarly the electrostatic potential $\overline{V}_{s} \left( z \right)$
 can be expressed within $R_{s}$ as a comparable Fourier series,

 \begin{equation}
   \overline{V}_{s} \left( z \right) = 
   \sum_{n} B^{(s)}_{n} e^{i \kappa^{(s)}_{n} z},
   \label{eq:ftpot}
 \end{equation}

 \noindent with the $B^{(s)}_{n}$ fixed by the Poisson equation, 
 Eq. (\ref{eq:transavpoisson}),

 \begin{equation}
    B^{(s)}_{n} = \frac{4 \pi A^{(s)}_{n}}
                  {\left[\kappa^{(s)}_{n}\right]^{2}}, \:\: n \ne 0,
   \label{eq:ftpotcoef}
 \end{equation}

 \noindent except for $B^{(s)}_{0}$ which is influenced by the charge
 density outside of $R^{'}_{s}$.
 Upon nanosmoothing, all contributions to 
 $\overline{\overline{V}}_{u} \left( z \right)$
 for $z$ within $R^{'}_{s}$ vanish except
 that for $n = 0$,

 \begin{equation}
    \overline{\overline{V}}_{s} \left( z \right) = B^{(s)}_{0}, 
   \label{eq:ftpotcoef0}
 \end{equation}

 \noindent which is invariant to the smoothing process.

 The potential shift $\Delta \overline{\overline{V}}_{u}$,
 defined in Eq. (\ref{eq:deltav}) as

 \begin{equation}
    \Delta \overline{\overline{V}}_{u} = 
    \overline{\overline{V}}_{u} \left(z_{2}\right) -
    \overline{\overline{V}}_{u} \left(z_{1}\right) ,
    \label{eq:diffdeltaV}
 \end{equation}

 \noindent is thus invariant to the smoothing procedure,

 \begin{equation}
    \Delta \overline{\overline{V}}_{u} = 
    B^{(2)}_{0} - B^{(1)}_{0}.
 \end{equation}

 \noindent Moreover, according to Eq. (\ref{eq:relshiftdip}),
 the nanosmoothed dipole-moment density $\overline{\overline{p}}_{u}$
 is invariant as well,

 \begin{equation}
    \overline{\overline{p}}_{u} = 
    \frac{1}{4 \pi} \Delta \overline{\overline{V}}_{u}  =
    \frac{1}{4 \pi} \left[ B^{(2)}_{0} - B^{(1)}_{0} \right].
 \end{equation}

 \subsection{The transferred charge density and the dipolar density.}
 \label{sec:nanosmoothing-transferch}

 It is of considerable physical interest to establish the value of the
 charge transferred across the interface, a difficult task.
 A criterion for establishing the position of the interface is needed,
 the difficulty of which is discussed in Sec. \ref{sec:refint}.
 The rapid, large oscillations of 
 $\overline{\rho}_{u} \left( z \right)$ and the relative
 smallness of the pertinent component of $\overline{\rho}_{u} \left( z \right)$ 
 make using it impractical.
 On the other hand, if one uses a criterion based on
 $\overline{\overline{\rho}}_{u} \left( z \right)$,
 $z_{int}$ can be sensitive to the smoothing function.
 Nevertheless, we shall attack the problem
 using $\overline{\overline{\rho}}_{u} \left( z \right)$ and attempt
 to overcome the resulting sensitivity to the smoothing function
 of the position of $z_{int}$ and the amount of charge transferred.

 We start by defining two cumulative charge densities 

 \begin{subequations}
    \begin{align}
       \overline{\overline{Q}}_{-} \left( z \right) & =
       \int_{z_{1}}^{z} dz^{'} \overline{\overline{\rho}}_{u} ( z^{'} ), 
       \label{eq:cumul-}
       \\
       \overline{\overline{Q}}_{+} \left( z \right) & =
       \int_{z}^{z_{2}} dz^{'} \overline{\overline{\rho}}_{u} ( z^{'} ).
       \label{eq:cumul+}
    \end{align}
 \end{subequations}

 For the unpolarized case now under consideration,
 $\overline{\overline{Q}}_{u} = 0$. Thus, as

 \begin{align}
    & \overline{\overline{Q}}_{u} = 
      \overline{\overline{Q}}_{-} \left( z \right) +
      \overline{\overline{Q}}_{+} \left( z \right), \:\:\: z \in (z_{1},z_{2}),
    \label{eq:quqpm} \\
    & \overline{\overline{Q}}_{-} \left( z \right) = 
      - \overline{\overline{Q}}_{+} \left( z \right)
 \end{align}

 \noindent holds for $\forall z \in (z_{1},z_{2})$, and 
 it is sufficient to consider either one or the other.
 Define $q$ as the magnitude of the charge transferred 
 per unit area of the interface, the transferred charge density.
 We estimate $q$ as 

 \begin{equation}
  q =\sup_{z} \left| \overline{\overline{Q}}_{\pm} \left( z \right)\right|,
  \label{eq:qchartrans}
 \end{equation}

 \noindent and estimate $z_{int}$ as 

 \begin{equation}
   z_{int} = arg \:\: 
   \sup_{z} \left| \overline{\overline{Q}}_{\pm} \left( z \right)\right|.
  \label{eq:defzint}
 \end{equation}

 Now both $q$ and $z_{int}$ are sensitive to the choices of $l_{1}$ and
 $l_{2}$ in the smoothing function $f$. 
 As $l_{1}$ and $l_{2}$ increase, $\omega_{l_{1}}$ can reach
 across the interface from material 1 to material 2
 bringing contributions from 
 $\overline{\rho}(z^{'})$, $z^{'}$ within material 2, to 
 $\overline{\overline{\rho}}(z)$, $z$ within material 1, and vice versa,
 thus returning part of the transferred charge back to its origin and reducing
 the value of $q$. 
 Moreover, since $\overline{\rho} \left( z \right)$ contains
 components which oscillate strongly with $z$,
 $\left| \overline{\overline{Q}}\left( z \right)\right|$ could 
 develope multiple suprema or maxima as $l_{1}$ and $l_{2}$ increase
 in multiples of the lattice constants of materials 1 and 2, respectively.
 This would vitiate the utility of the definitions (\ref{eq:qchartrans}) and
 (\ref{eq:defzint}) of the transferred charge $q$ and the 
 interface location $z_{int}$, respectively, should it happen.
 We have found that it does happen in the toy model described in
 Sec. \ref{sec:toymodel} 
 and studied in Sec. \ref{sec:resultsunpolar}, 
 cf. Fig. \ref{fig:diffr} below,  
 in the case where the interatomic distance remains unchanged
 across the entire superlattice. 
 Accordingly, as a precaution, the smallest 
 acceptable values of $l_{1}$ and $l_{2}$ 
 should be used for $f \left( z \right)$, a single lattice constant 
 of each material, an important additional condition on $l_{1}$ and $l_{2}$.
 
 If the transferred charge density were concentrated equally on two surfaces
 at either side of the interface, separated by a distance $\lambda$,
 a dipole moment density of magnitude $ q \lambda$ would be created.
 Setting $ q \lambda$  equal to the actual dipolar density 
 $\overline{\overline{p}}_{u}$ allows us to define $\lambda_{u}$
 as the dipolar length

 \begin{equation}
    \lambda_{u} = \frac{\overline{\overline{p}}_{u}}{q_{u}},
 \end{equation}

 \noindent where we have restored the subscript $u$ to $q$ as we are
 dealing with the unpolarized case. 

 \subsection{Loss of invariance of physical magnitudes of interest 
             with nanosmoothing.}
 \label{sec:looseinvariance}

 Now we can ask whether the physical magnitudes of interest, such as 
 the interfacial charge [Eq. (\ref{eq:qu})] or dipole moment 
 densities [Eq. (\ref{eq:pu})]
 remain unchanged if the microscopic charge density is replaced with 
 the nanosmoothed charge density. In other words,
 if we define the interfacial charge $\overline{\overline{Q}}_{u}$
 and dipolar densities $\overline{\overline{p}}_{u}$
 computed from $\overline{\overline{\rho}}_{u} \left( z \right)$ as

 \begin{subequations}
    \begin{align}
      \overline{\overline{Q}}_u & = \int_{z_{1}}^{z_{2}} dz \:\: 
      \overline{\overline{\rho}}_u \left( z \right), 
      \label{eq:bbqu} 
      \\
      \overline{\overline{p}}_u & = \int_{z_{1}}^{z_{2}} dz \:\: z 
      \overline{\overline{\rho}}_u \left( z \right), 
      \label{eq:bbpu}
    \end{align}
 \end{subequations}

 \noindent then the question is whether 
 $\overline{\overline{Q}}_u = Q_{u}$ and 
 $\overline{\overline{p}}_u = p_{u}$ for given integration limits
 $z_{1}$ and $z_{2}$. 

 Replacing Eq. (\ref{eq:bbrho}) into Eq. (\ref{eq:bbqu}),

 \begin{equation}
    \overline{\overline{Q}}_u  = \int_{z_{1}}^{z_{2}} dz \:\: 
    \overline{\overline{\rho}}_u \left( z \right) =
    \int_{z_{1}}^{z_{2}} dz \:\: 
    \int_{z-L}^{z+L} d z^{'} f ( z - z^{'} )
    \overline{\rho}_{u} ( z^{'} ).
    \label{eq:bbqu1}
 \end{equation}

 \begin{figure}[htbp]
    \begin{center}
       \includegraphics[width=6cm,angle=0] {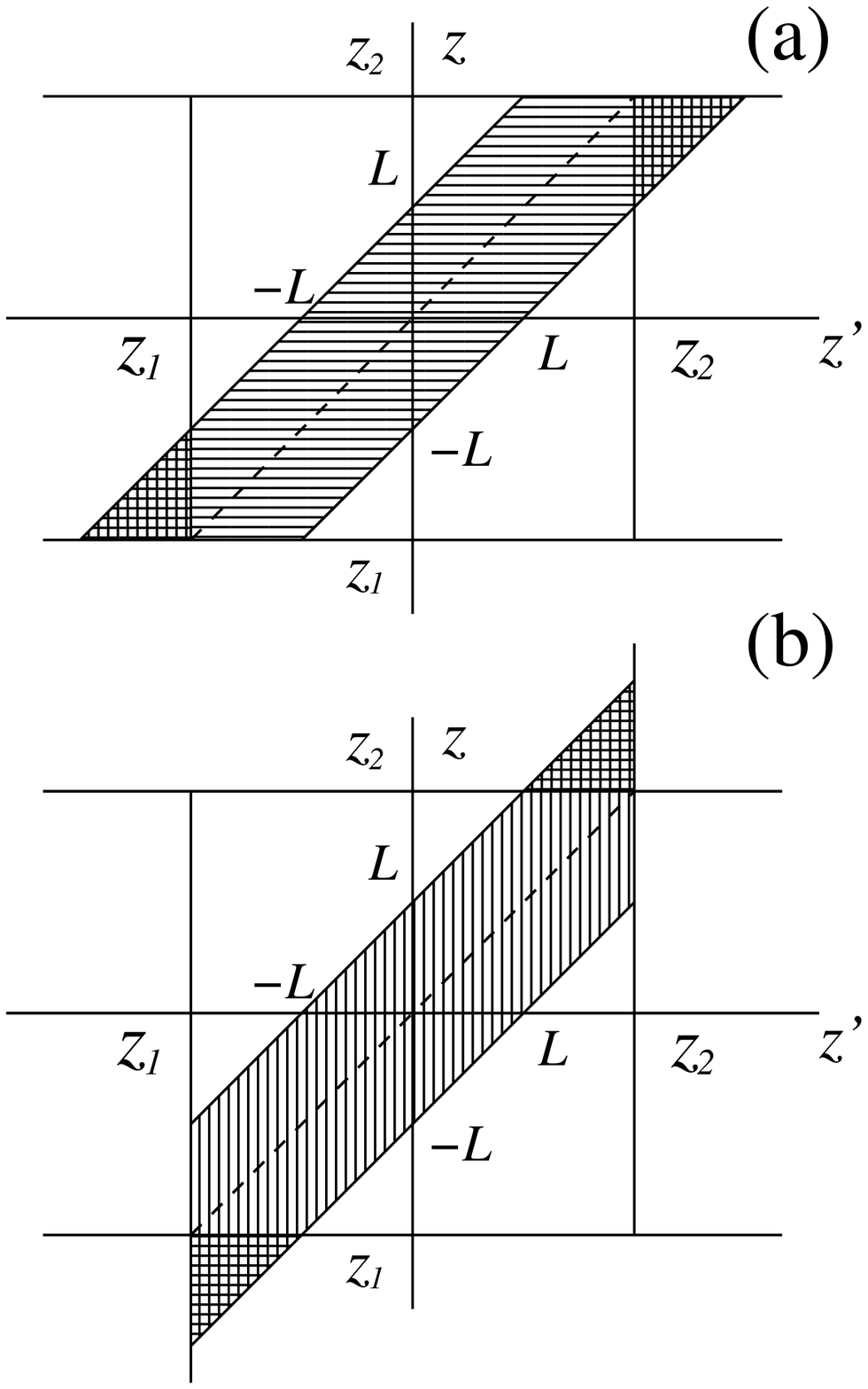}
       \caption{ (a) The range of integration in the $z$, $z^{'}$ plane
                 according to Eq. (\ref{eq:bbqu1}).
                 (b) The range of integration according to
                 Eq. (\ref{eq:qu1}).
                 The direction of the shading lines, horizontal in (a)
                 and vertical in (b), indicates the variable of which
                 the first integral occurs, over $z^{'}$ in
                 Eq. (\ref{eq:bbqu1}), and $z$ in Eq. (\ref{eq:qu1})
                 respectively.
               }
       \label{fig:rangeint}
    \end{center}
 \end{figure}

 \noindent The region of integration in Eq. (\ref{eq:bbqu1}) within the
 $z$, $z^{'}$ plane is shaded in Fig. \ref{fig:rangeint}(a). Then, the integral 
 of Eq. (\ref{eq:bbqu1}) can be decomposed into the integral on the central 
 square plus the integrals on the cross hatched triangles 

 \begin{align}
    \overline{\overline{Q}}_u  = &  
         \int_{z_{1}}^{z_{2}} dz \:\: 
         \int_{z_{1}}^{z_{2}} dz^{'} \:\: 
          f ( z - z^{'} ) \overline{\rho}_{u} ( z^{'} ) +
    \nonumber \\
         & \int_{z_{2}-L}^{z_{2}} dz \:\: 
         \int_{z_{2}}^{z+L} dz^{'} \:\: 
          f ( z - z^{'} ) \overline{\rho}_{u} ( z^{'} ) +
    \nonumber \\
         & \int_{z_{1}}^{z_{1}+L} dz \:\: 
         \int_{z-L}^{z_{1}} dz^{'} \:\: 
          f ( z - z^{'} ) \overline{\rho}_{u} ( z^{'} ).
    \label{eq:bbqu2}
 \end{align}

 On the other hand, $Q$ is defined as

 \begin{equation}
    Q = \int_{z_{1}}^{z_{2}} dz^{'} \:\:
        \overline{\rho}_{u} ( z^{'} )  =
        \int_{z_{1}}^{z_{2}} dz^{'} \:\:
        \overline{\rho}_{u} ( z^{'} )
        \int_{z^{'}-L}^{z^{'}+L} \:\: dz f ( z - z^{'} ),
        \label{eq:qu1}
 \end{equation}

 \noindent where the region of integration in the
 $z$, $z^{'}$ plane is now shaded in Fig. \ref{fig:rangeint}(b). 
 Decomposing the domain of integration as for 
 $\overline{\overline{Q}}_{u}$ above gives 

 \begin{align}
    Q_u  = &  
         \int_{z_{1}}^{z_{2}} dz \:\: 
         \int_{z_{1}}^{z_{2}} dz^{'} \:\: 
          f ( z - z^{'} ) \overline{\rho}_{u} ( z^{'} ) +
    \nonumber \\
         & \int_{z_{2}}^{z_{2}+L} dz \:\: 
         \int_{z-L}^{z_{2}} dz^{'} \:\: 
          f ( z - z^{'} ) \overline{\rho}_{u} ( z^{'} ) +
    \nonumber \\
         & \int_{z_{1}-L}^{z_{1}} dz \:\: 
         \int_{z_{1}}^{z+L} dz^{'} \:\: 
          f ( z - z^{'} ) \overline{\rho}_{u} ( z^{'} ).
    \label{eq:qu2}
 \end{align}

 Therefore, the equality $Q_{u} = \overline{\overline{Q}}_{u}$ is verified
 if and only if
 the integrals over the shaded regions on either side of the line $z = z^{'}$
 are the same, that is if the following pair of equations hold

 \begin{subequations}
    \begin{align}
       \int_{z_{2}-L}^{z_{2}} dz \:\: 
       \int_{z_{2}}^{z+L} dz^{'} \:\: 
         f ( z - z^{'} ) \overline{\rho}_{u} ( z^{'} ) & = 
       \int_{z_{2}}^{z_{2}+L} dz \:\: 
       \int_{z-L}^{z_{2}} dz^{'} \:\: 
         f ( z - z^{'} ) \overline{\rho}_{u} ( z^{'} ) ,
    \label{eq:condtriangle1} \\
       \int_{z_{1}}^{z_{1}+L} dz \:\: 
       \int_{z-L}^{z_{1}} dz^{'} \:\: 
          f ( z - z^{'} ) \overline{\rho}_{u} ( z^{'} ) & =
       \int_{z_{1}-L}^{z_{1}} dz \:\: 
       \int_{z_{1}}^{z+L} dz^{'} \:\: 
          f ( z - z^{'} ) \overline{\rho}_{u} ( z^{'} ).
    \label{eq:condtriangle2} 
    \end{align}
 \end{subequations}

 \noindent If in Eq. (\ref{eq:condtriangle1}) we apply the following change of
 variables $z = z_{2} - u$ and $z^{'} = z_{2} + v$ in the left hand side 
 and $z = z_{2}+u$ and $z^{'} = z_{2} - v$ in the right hand side, 
 then Eq. (\ref{eq:condtriangle1}) transforms into

 \begin{equation}
    \int_{0}^{L} du \:\: 
    \int_{0}^{L-u} dv \:\: 
      f ( -u-v ) \:\:  \overline{\rho}_{u} ( z_{2}+v )  = 
    \int_{0}^{L} du \:\: 
    \int_{0}^{L-u} dv \:\: 
      f ( u+v ) \:\: \overline{\rho}_{u} ( z_{2}-v ) .
 \end{equation}

 \noindent Due to the parity conditions of the filter function, we know that
 $f( -u-v ) = f ( u+v ) $. 
 But, $\overline{\rho}_{u} ( z_{2}+v ) = \overline{\rho}_{u} ( z_{2}-v )$
 if and only if $\overline{\rho}$ is even about $z_{2}$.
 Completely analogous 
 reasoning can be applied to Eq. (\ref{eq:condtriangle2}) and is 
 omitted here. 
 Thus, for $\overline{\overline{Q}}_{u}$ to equal $Q_{u}$, it must be possible
 to find a $z_{1}$ and a $z_{2}$ about which $\overline{\rho}(z)$ 
 is {\it symmetric} for $\left| z - z_{1,2}\right| \le L$.

 We now show for the unpolarized case that this symmetry condition can
 be satisfied. For a multilattice consisting on alternating layers
 of two different materials $s$, $s = 1$ or $2$, there are two
 distinct interfaces within the supercell bounded by $z_{1}$ and
 $z_{1}+a_{sc}$, interface 1,2 and interface 2,1.
 We define total charge $Q_{sc}$ and $\overline{\overline{Q}}_{sc}$
 which are the sums of the charges associated with each individual interface,

 \begin{subequations}
    \begin{align}
      Q_{sc} & = Q_{12} + Q_{21} = \int_{z_{1}}^{z_{1}+a_{sc}} dz^{'} 
                 \:\: \overline{\rho}_u ( z^{'} ), 
      \label{eq:qsc} 
      \\
      \overline{\overline{Q}}_{sc} & = 
      \overline{\overline{Q}}_{12} + 
      \overline{\overline{Q}}_{21} = 
      \int_{z_{1}}^{z_{1}+a_{sc}} dz
                 \:\: \overline{\overline{\rho}}_u ( z ).
      \label{eq:qscbar}
    \end{align}
 \end{subequations}

 \noindent Proceeding in analogy with Eq. (\ref{eq:bbqu2}) and 
 Eq. (\ref{eq:qu2}), we obtain 

 \begin{align}
    \overline{\overline{Q}}_{sc}  = &  
         \int_{z_{1}}^{z_{1}+a_{sc}} dz \:\: 
         \int_{z_{1}}^{z_{1}+a_{sc}} dz^{'} \:\: 
          f ( z - z^{'} ) \overline{\rho}_{u} ( z^{'} ) +
    \nonumber \\
         & \int_{z_{1}+a_{sc}-L}^{z_{1}+a_{sc}} dz \:\: 
         \int_{z_{1}+a_{sc}}^{z+L} dz^{'} \:\: 
          f ( z - z^{'} ) \overline{\rho}_{u} ( z^{'} ) +
    \nonumber \\
         & \int_{z_{1}}^{z_{1}+L} dz \:\: 
         \int_{z-L}^{z_{1}} dz^{'} \:\: 
          f ( z - z^{'} ) \overline{\rho}_{u} ( z^{'} ),
    \label{eq:qsc2}
 \end{align}

 \begin{align}
    Q_{sc}  = &  
         \int_{z_{1}}^{z_{1}+a_{sc}} dz \:\: 
         \int_{z_{1}}^{z_{1}+a_{sc}} dz^{'} \:\: 
          f ( z - z^{'} ) \overline{\rho}_{u} ( z^{'} ) +
    \nonumber \\
         & \int_{z_{1}+a_{sc}}^{z_{1}+a_{sc}+L} dz \:\: 
         \int_{z-L}^{z_{1}+a_{sc}} dz^{'} \:\: 
          f ( z - z^{'} ) \overline{\rho}_{u} ( z^{'} ) +
    \nonumber \\
         & \int_{z_{1}-L}^{z_{1}} dz \:\: 
         \int_{z_{1}}^{z+L} dz^{'} \:\: 
          f ( z - z^{'} ) \overline{\rho}_{u} ( z^{'} ).
    \label{eq:qsc3}
 \end{align}

 The fact that $\overline{\rho}_{u} (z) $ is periodic in $z$
 with period $a_{sc}$ allows as to rewrite 
 Eqs. (\ref{eq:qsc2}) and (\ref{eq:qsc3})
 so as to establish the equality of $Q_{sc}$ and 
 $\overline{\overline{Q}}_{sc}$,

 \begin{align}
    \overline{\overline{Q}}_{sc}  = &  
         \int_{z_{1}}^{z_{1}+a_{sc}} dz \:\: 
         \int_{z_{1}}^{z_{1}+a_{sc}} dz^{'} \:\: 
          f ( z - z^{'} ) \overline{\rho}_{u} ( z^{'} ) +
    \nonumber \\
         & \int_{z_{1}-L}^{z_{1}} dz \:\: 
         \int_{z_{1}}^{z+L} dz^{'} \:\: 
          f ( z - z^{'} ) \overline{\rho}_{u} ( z^{'} ) +
    \nonumber \\
         & \int_{z_{1}}^{z_{1}+L} dz \:\: 
         \int_{z-L}^{z_{1}} dz^{'} \:\: 
          f ( z - z^{'} ) \overline{\rho}_{u} ( z^{'} ) = Q_{sc}
    \label{eq:qsc4}
 \end{align}

 Since the supercell is electrically neutral, so must the 
 nanosmoothed supercell be,

 \begin{equation}
    Q_{sc} = 0 = \overline{\overline{Q}}_{sc}.
 \end{equation}

 \noindent Consequently, from Eq. (\ref{eq:qscbar}) it follows that

 \begin{equation}
    \overline{\overline{Q}}_{12} = -\overline{\overline{Q}}_{21},
 \end{equation}

 \noindent implying that there would be a smooth electrostatic field
 within each layer if $\overline{\overline{Q}}_{12}$ is nonzero.
 The existence of such a field would polarize the system in 
 contradiction to the initial condition that the system is unpolarized.
 We conclude that 

 \begin{equation}
    \overline{\overline{Q}}_{12} = \overline{\overline{Q}}_{21} = 0.
 \end{equation}

 Thus, for the interface charge to be invariant to nanosmoothing,
 that is, for $\overline{\overline{Q}}_{12} = Q_{12}$
 to hold, $z_{1}$ and $z_{2}$ must be positioned in $R_{1}$ and $R_{2}$
 so that $Q_{12}$ vanishes in the unpolarized case. 
 To do this, one could make an arbitrary choice
 of $z_{1}$ in $R_{1}$, say, and then integrate 
 $\overline{\rho}_{u} (z)$ from $z_{1}$ up to some $z_{2}$ in $R_{2}$
 at which the integral vanishes. There is no need to do this, as it is
 the nanosmoothing quantities themselves which are of interest.

 Repeating the reasoning for the dipole moment density, we arrive at
 the conclusion that for $\overline{\overline{p}}_{u} = p_{u}$,
 the following condition must be satisfied

 \begin{equation}
    \int_{z_{2}-L}^{z_{2}} dz \:\: z 
    \int_{z_{2}}^{z+L} dz^{'} \:\: 
      f ( z - z^{'} ) \overline{\rho}_{u} ( z^{'} )  = 
    \int_{z_{2}}^{z_{2}+L} dz \:\: 
    \int_{z-L}^{z_{2}} dz^{'} \:\: z^{'} 
      f ( z - z^{'} ) \overline{\rho}_{u} ( z^{'} ) ,
    \label{eq:condtrianglep1}
 \end{equation}

 \noindent Applying the same change of variables as before, 
 that is
 $z = z_{2} - u$ and $z^{'} = z_{2} + v$ in the left hand side 
 and $z = z_{2}+u$ and $z^{'} = z_{2} - v$ in the right hand side, 
 then Eq. (\ref{eq:condtrianglep1}) transforms into

 \begin{equation}
    \int_{0}^{L} du \:\:\: (z_{2}-u)
    \int_{0}^{L-u} dv \:\:\: 
      f ( -u-v ) \:\: \overline{\rho}_{u} ( z_{2}+v )  = 
    \int_{0}^{L} du \:\: 
    \int_{0}^{L-u} dv \:\:\: 
      f ( u+v ) \:\: \overline{\rho}_{u} ( z_{2}-v ) \:\: (z_{2}-v).
 \end{equation}

 \noindent Even in the case of a function $\overline{\rho}_{u}$ symmetric
 around $z_{1}$ and $z_{2}$, the previous condition does not hold in general,
 and the difference between $\overline{\overline{p}}_{u}$ and $p_{u}$
 amounts to 

 \begin{equation}
    \int_{0}^{L} du \:\: 
    \int_{0}^{L-u} dv \:\: (u-v) \:\:
      f ( u+v ) \:\: \overline{\rho}_{u} ( z_{2}+v ) ,  
 \end{equation}

 \noindent plus a similar term that comes from the difference in
 the lower triangles in Fig. \ref{fig:rangeint}(a) and 
 Fig. \ref{fig:rangeint}(b).
 This has to be evaluated for each particular case.

 In conclusion, the interface dipole-moment density is not invariant to
 nanosmoothing and the interface charge density can be made so only
 by exquisite case in the choice of $z_{1}$ and $z_{2}$.
 This is of no concern, as it is the nanosmoothed quantites which 
 are of physical interest.

\section{Description of the toy model.}
\label{sec:toymodel}

 As highlighted in the introduction, the components of the density
 which give rise to the interface-related
 dipole densities are nearly obscured by the much larger variations
 of the total microscopic charge density.
 This atomic-scale charge density, routinely provided by
 any density-functional-based first-principles code, 
 is affected by numerical noise and convergence problems
 inherent in some of the standard approximations in the
 practical implementations of density functional theory (DFT). 
 Therefore, the accuracy of the computations required for extracting the 
 actual charge transferred from one side of the interface
 to the other must be high enough so that the numerical noise
 of the calculations is orders of magnitude smaller than 
 the relevant interface-related charge densities.
 In this paper, in order to illustrate all of the essential 
 elements of the 
 theory of smoothing while avoiding these practical problems,
 we shall define a toy interface model that resembles
 closely a realistic multilayer material but
 whose computational accuracy can be systematically improved.

 The requirements that such a toy model should meet are:
 (i) its electron density  must be a continuous function;
 (ii) far away from the interfaces, where the interface-induced perturbation
 of the charge density becomes negligible, 
 the toy electron density must tend to two distinct periodic functions 
 on the left and on the right of each interface, 
 mimicking the differing behaviour at the bulk level of the materials that
 constitute the multilayer system containing the interfaces;
 and (iii) the interlayer spacing at the interfaces should be distorted
 with respect those at bulk so as to simulate better the interface
 induced relaxations that happen in real interfaces.

 In the toy model we propose here, we represent only the laterally averaged
 density, a one-dimensional function.
 As before, the direction perpendicular to the interface
 is referred to as the $z$ axis.

 We shall consider atomic-like charge densities $g_{i}^{(s)}$ 
 (again as before, the superindex $s= \left\{ 1,2 \right\}$ refers to the side
 of the interface, left or right, where a given ``atom'' $i$ is located)
 of the sum of two gaussians, 
 centered at each atomic site $z_{i}^{(s)}$,  

 \begin{align}
    g^{(s)}_{i} \left( z - z^{(s)}_{i} \right) & = 
     \frac{A^{n,s}_{i}} {\sigma_{n,s} \sqrt{2\pi}} 
     exp\left[ - \frac{\left(z-z^{(s)}_{i}\right)^{2}}{2 \sigma_{n,s}^{2}} \right]
    \nonumber \\
    & - \frac{A^{e,s}_{i}} {\sigma_{e,s} \sqrt{2\pi}} 
    exp\left[ - \frac{\left(z-z^{(s)}_{i}-\delta^{(s)}_{i}\right)^{2}}
    {2 \sigma_{e,s}^{2}} \right].
 \end{align}

 \noindent The positive (negative) gaussian, 
 whose standard deviation is denoted by $\sigma_{n}$ 
 ($\sigma_{e}$), mimics the nuclear charge
 density (electronic charge density) 
 of atom $i$ centered on position $z_{i}$.
 By imposing $\sigma_{n} < \sigma_{e}$ we make
 the ``nuclear'' charge more confined than the ``electronic'' charge.
 Since the gaussians are normalized, a net charge per atomic site
 can be simulated by making $A^{n,s}_{i} \ne A^{e,s}_{i}$.
 The parameter $\delta^{(s)}_{i}$ allows us to displace the
 electronic clouds with respect to the nuclei and thereby produce
 a net dipole moment on a particular atom.

 The atomic-like charge densities are arranged in bulk unit cells
 that might contain one single atom or a more complicated
 polyatomic basis. 
 In this Section we shall assume that the bulk unit cells of each 
 of the materials
 that form the superlattice contains a single atom that does not carry
 any charge ($A_{i}^{n,s} = A_{i}^{e,s} \:\: \forall \:\: i$) or dipole moment
 ($\delta^{(s)}_{i} = 0 \:\: \forall \:\: i$). 
 For the polarized case, we refer the reader
 to Sec. \ref{sec:resultspolar}.
 Thus, for the unpolarized interface each material can be considered 
 as a one-dimensional monoatomic chain, where consecutive ``atoms'' 
 are separated by a distance $a^{(s)}$.
 The interatomic distance at the interface, $a_{int}$, is taken as 
 $a_{int} = \frac{ a^{(1)} + a^{(2)} }{2}$. 
 Then, consecutive interatomic distances from the interface evolve smoothly 
 towards the bulk value as a function of the distance to the interface.
 In our simulations, when we move from the interface towads material 
 $s, s= \left\{ 1,2 \right\},$ 
 the second interatomic distance is set up to 
 $\frac{1}{4} a^{(o)} + \frac{3}{4} a^{(s)}$,
 where $a^{(o)}$ is the lattice constant of the other material.
 The bulk value is recovered only at the third interatomic distance
 from the surface.

 A supercell is then built, as described in Sec. \ref{sec:first-principles}.
 The basic unit cell,  periodically repeated
 in space, contains a suitable number $N_{1}$ and $N_{2}$ 
 of bulk unit cells of the two materials.
 The microscopic charge density, 
 $\overline{\rho} \left( z \right)$, is 
 defined as the superposition of all the atomic-like charge densities

 \begin{align}
    \overline{\rho}\left( z \right) = 
    \sum_{i=1}^{N_{1}} g_{i}^{(1)} ( z - z_{i}^{(1)} )
    + \sum_{j=1}^{N_{2}} g_{j}^{(2)} ( z - z_{j}^{(2)} ).
    \label{eq:combgauss}
 \end{align}

\begin{figure}[htbp]
    \begin{center}
       \includegraphics[width=10cm] {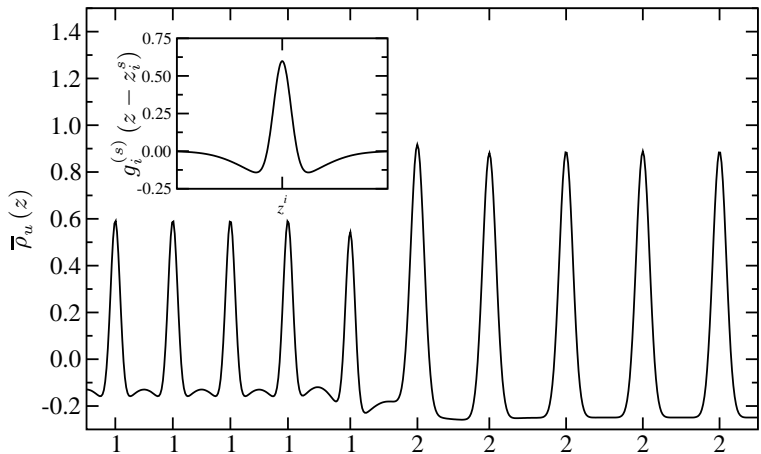}
       \caption{ Illustration of the toy model used to simulate 
         microscopic charge-densities in a multilayered material.
         A non-polar case is represented ($\delta^{(s)}_{i} = 0$
	 and $A^{n,s}_{i} = A^{e,s}_{i} \:\: \forall \:\: i$).
         The remaining parameters have been chosen as 
         $A^{n,1} = A^{e,1} = 1$, 
         $\sigma_{n,1} = 0.5$,
         $\sigma_{e,1} = 2.0$,
         $A^{n,2} = A^{e,2} = 2$, 
         $\sigma_{n,2} = 0.7$,
         $\sigma_{e,2} = 4.5$, 
         $\delta^{(1)} = \delta^{(2)} = 0$ for all the ``atoms'', 
	 $a^{(1)} = 6$, and  
	 $a^{(2)} = 8$.
         Within each period of the multilayer, there are 10 atomic layers
         of the left material and 10 of the right.
         Only a small portion, centered at the interface,
         of the microscopic charge density of the supercell is shown here.
         Inset: charge density of an individual ``atom'' on the left. 
         Atomic units are used.
         }
       \label{fig:nanofun}
    \end{center}
 \end{figure}

 \noindent The resulting model is illustrated for the non polar case 
 ($\delta^{(s)}_{i} = 0 $, and $A^{n,s}_{i} = A^{e,s}_{i} \:\: \forall \:\: i$) 
 in Fig. \ref{fig:nanofun};
 the parameters of the model are specified in the figure caption.
 An isolated atomic-like charge density 
 is shown in the inset.
 The sizes of the layers of the two materials 
 that constitute the multilayer can 
 be tuned by changing the number of building blocks in the left, $N_{1}$,
 or in the right, $N_{2}$.

 $\overline{\overline{\rho}} \left( z \right)$ is formed from 
 $\overline{\rho} \left( z \right)$ by convoluting it with the
 filter function. Such convolutions are most conveniently formed 
 by fast Fourier transforms, which require the discretization of
 space into a uniformly-spaced grid of points.
 
 \section{Results: toy model, non polar case.}
 \label{sec:resultsunpolar}

 \psfrag{R1}[cc][cc]{$R_{1}$}
 \psfrag{R2}[cc][cc]{$R_{2}$}
 \psfrag{Rp1}[cc][cc]{$R^{'}_{1}$}
 \psfrag{Rp2}[cc][cc]{$R^{'}_{2}$}

 \begin{figure}[htbp]
    \begin{center}
       \includegraphics[width=10cm]{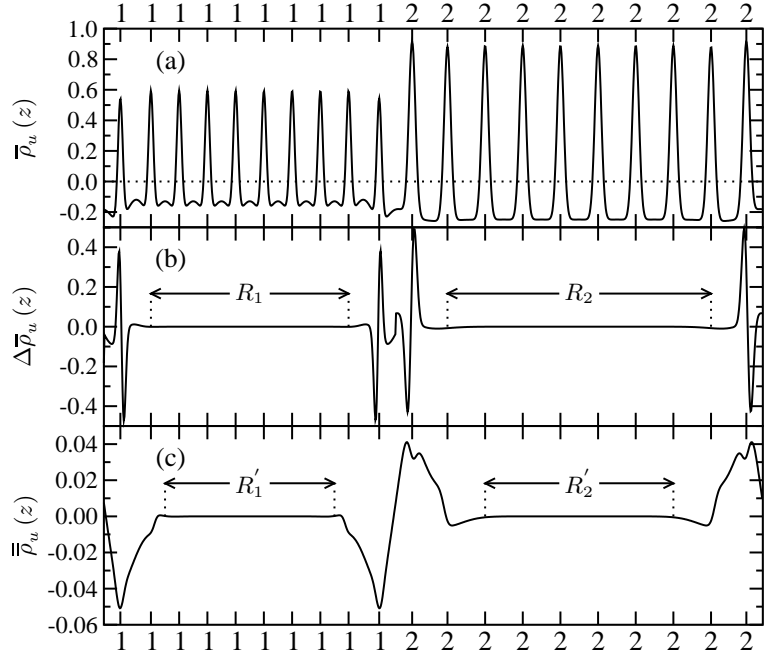}
       \caption{ (a) Microscopic charge density 
                 $\overline{\rho}_{u} \left( z \right)$ within a unit supercell
                 for an unpolarized system.
                 The widths of the left (1) and right (2) materials
                 have both been set to 10 atomic layers, with 
                 interatomic separations of $a^{(1)} = 6$ and 
		 $a^{(2)} = 8$. The rest of the
                 parameters of the toy model are as in the caption of Fig.
                 \ref{fig:nanofun}.
                 (b) Interface-induced charge density 
                 $\Delta \overline{\rho}_{u} \left( z \right)$ 
                 defined by subtracting 
                 the charge density of a reference system,
                 defined as in Eq. (\ref{eq:rho0}).
                 The interface plane $z_{int}$ is positioned at
                 the middle of the
                 separation between the last atomic plane on the
                 left and the first atomic plane on the right.
                 (c) Nanosmoothed charge density 
                 $\overline{\overline{\rho}}_{u}(z)$ of the microscopic charge 
                 density. The nanosmoothing function is defined as in 
                 Eq. (\ref{eq:filterconvolution}) with 
                 $l_{1} = a^{(1)} = 6$, and 
		 $l_{2} = a^{(2)} = 8$,  single interplanar 
                 distances along $z$ for each material.
                 Atomic units are used.
         }
       \label{fig:nanotoy}
    \end{center}
 \end{figure}

 In Fig. \ref{fig:nanotoy}(a) we illustrate the specific toy model we shall 
 analyze in detail for the unpolarized case. The parameters of the model are
 $A^{n,1} = A^{e,1} = 1$, 
 $\sigma_{n,1} = 0.5$,
 $\sigma_{e,1} = 2.0$,
 $A^{n,2} = A^{e,2} = 2$, 
 $\sigma_{n,2} = 0.7$,
 $\sigma_{e,2} = 4.5$, 
 $\delta^{(1)} = \delta^{(2)} = 0$ for all the ``atoms'', 
 $a^{(1)} = 6$, and  
 $a^{(2)} = 8$.
 Atomic units are used throughout the paper.
 First, we consider a reference density 
 $\overline{\rho}_{0} \left( z \right)$ defined as
 in Eq. (\ref{eq:rho0}), locating the interface at $z_{int}$ half way 
 between the rightmost atomic layer of material 1 
 and the leftmost atomic layer of material 2.
 $\Delta \overline{\rho}_{u} \left( z \right)$, 
 defined as in Eq. (\ref{eq:deltarho}),
 is plotted in Fig. \ref{fig:nanotoy}(b). Next, we
 construct $\overline{\overline{\rho}}_{u} \left( z \right)$ 
 from $\overline{\rho}_{u} \left( z \right)$ directly 
 using a nanosmoothing function $f (z - z^{'})$ 
 defined as in Eq. (\ref{eq:filterconvolution}),
 in which $l_{1}=a^{(1)}=6$, and $l_{2}=a^{(2)}=8$.
 The smoothed charge density 
 $\overline{\overline{\rho}}_{u} \left( z \right)$ 
 is shown in Fig. \ref{fig:nanotoy}(c).
 As we have already discussed in Sec. \ref{sec:nanosmoothing-insensitivity},
 the regions $R^{'}_{s}$ within which $\overline{\overline{\rho}}_{u}$
 vanishes lie within the regions $R_{s}$ within which 
 $\Delta \overline{\rho}_{u}$ vanishes, 
 because smoothing $\overline{\rho}_{u}$ 
 within $R_{s}$ brings into $\overline{\overline{\rho}}_{u}$
 values of $\overline{\rho}_{u}$ for $z$ outside $R_{s}$.
 Due to the symmetry of the microscopic charge density at the center
 of each layer, the interfacial charge density 
 between material 1 and material 2 is the mirror image of the 
 interfacial charge density between material 2 and material 1
 [see the center and the edges of the Fig. \ref{fig:nanotoy}(b)
 and Fig. \ref{fig:nanotoy}(c)].

\begin{figure}[htbp]
    \begin{center}
       \includegraphics[width=10cm] {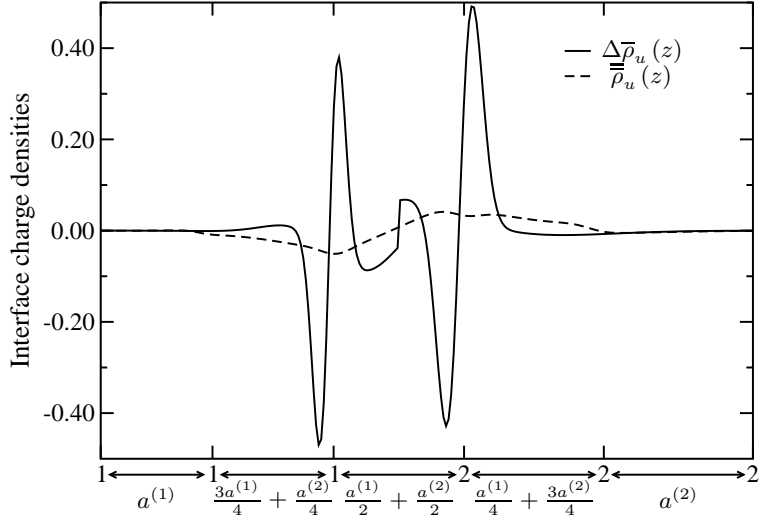}
       \caption{ Blow up of the interface-induced charge densities
                 obtained from the subtraction of the reference bulk 
                 density and from nanosmoothing. 
		 Their difference is clearly shown.
         }
       \label{fig:nanotoy-zoom}
    \end{center}
 \end{figure}

 A closer look at the interface region is shown in Fig. \ref{fig:nanotoy-zoom}.
 The discontinuity of $\Delta \overline{\rho}_{u}$ at the interface plane
 is clearly observed. Also, we can see how $\Delta \overline{\rho}_{u}$
 displays large fluctuations at the atomic scale in the neighborhood
 of the interface due to the interface-induced relaxations of the atomic
 layers. The positions of the atoms in the layers close to the
 interface do not coincide with the positions of the atoms after the
 cleaving of the bulk to define $\overline{\rho}_{0}$ 
 (see Sec. \ref{sec:refint}).
 Therefore, in the computation of $\Delta \overline{\rho}_{u}$ we are
 subtracting charge densities centered on different positions. 
 The nanosmoothing procedure eliminates not only the contributions
 associated with $\overline{\rho}_{0}$ in the bulk
 regions $R_{1}$ and $R_{2}$, but it filters out the oscillations
 due to the interface induced relaxations while producing a continuous
 charge distribution.
 Nanosmoothing $\overline{\rho}_{u}$ is clearly superior to forming 
 $\Delta \overline{\rho}_{u}$.

 \psfrag{d1}[cc][cc]{$a^{(1)}$}
 \psfrag{d2}[cc][cc]{$\frac{3a^{(1)}}{4}+\frac{a^{(2)}}{4}$}
 \psfrag{d3}[cc][cc]{$\frac{a^{(1)}}{2}+\frac{a^{(2)}}{2}$}
 \psfrag{d4}[cc][cc]{$\frac{a^{(1)}}{4}+\frac{3a^{(2)}}{4}$}
 \psfrag{d5}[cc][cc]{$a^{(2)}$}

\begin{figure}[htbp]
    \begin{center}
       \includegraphics[width=10cm]{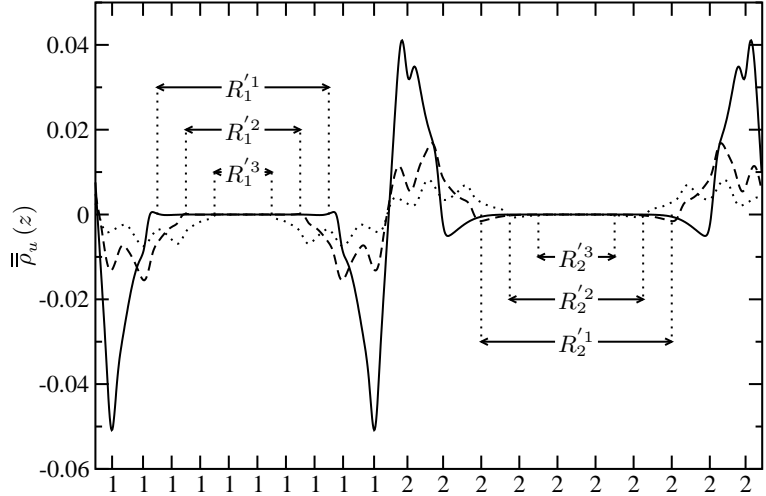}
       \caption{ Nanosmoothed charge density 
                 $\overline{\overline{\rho}}_{u} \left( z \right)$
                 obtained by nanosmoothing the planar average charge
                 density $\overline{\rho}_{u} \left( z \right)$ 
                 shown in Fig. \ref{fig:nanotoy}(a) with three different
                 filter functions. 
                 They are defined following 
                 Eq. (\ref{eq:filterconvolution}) with
                 $l_{1} = a^{(1)} = 6$, and $l_{2} = a^{(2)} = 8$ (solid line); 
                 $l_{1} = 2a^{(1)} = 12$, and $l_{2} = 2a^{(2)} = 16$ 
		 (dashed line); 
                 and 
                 $l_{1} = 3a^{(1)} = 18$, and $l_{2} = 3a^{(2)} = 24$ 
		 (dotted line),
                 respectively.
         }
       \label{fig:diffr}
    \end{center}
 \end{figure}

 In Fig. \ref{fig:diffr}, we show 
 $\overline{\overline{\rho}}_{u} \left( z \right)$ 
 for three different smoothing functions, defined as in 
 Eq. (\ref{eq:filterconvolution}) with
 $l_{1} = a^{(1)} = 6$, and $l_{2} = a^{(2)} = 8$ ; 
 $l_{1} = 2a^{(1)} = 12$, and $l_{2} = 2a^{(2)} = 16$ ; 
 and 
 $l_{1} = 3a^{(1)} = 18$, and $l_{2} = 3a^{(2)} = 24$, 
 respectively.
 The widths of the regions where 
 $\overline{\overline{\rho}}_{u} \left( z \right)$
 vanishes ($R^{'}_{1}$ and $R^{'}_{2}$) depend on the
 width of the filter function. As a general rule, 
 the more extended the filter function, the narrower the 
 $R^{'}$ regions.
 Although the profiles of the charge density differ
 significantly, the three charge distributions have the
 {\it same} net charge density $\overline{\overline{Q}}_{u}$ and
 dipole moment density $\overline{\overline{p}}_{u}$
 (see Table \ref{table:results-unpol}).
 This striking sensitivity of the nanosmoothed charge density
 to the smoothing function impedes detailed physical
 interpretation of its features.

 \psfrag{R11}[cc][cc]{$R^{'1}_{1}$}
 \psfrag{R12}[cc][cc]{$R^{'2}_{1}$}
 \psfrag{R13}[cc][cc]{$R^{'3}_{1}$}
 \psfrag{R21}[cc][cc]{$R^{'1}_{2}$}
 \psfrag{R22}[cc][cc]{$R^{'2}_{2}$}
 \psfrag{R23}[cc][cc]{$R^{'3}_{2}$}

 Even though we are dealing with an unpolarized interface made of the 
 juxtaposition of neutral and non-polar atoms, 
 the interface charge density $Q_{u}$,
 Eq. (\ref{eq:qu}), vanishes if and only if the integration limits
 $z_{1}$ and $z_{2}$ are taken midway between atoms inside each material
 layer, meeting the requirements for symmetry described in 
 Sec. \ref{sec:looseinvariance}
 (see Table \ref{table:results-unpol}). 
 $\Delta Q_{u}$, Eq. (\ref{eq:deltaq}), on the other hand, 
 does not vanish since, with our criterion for locating the interface
 plane $z_{int}$, we do not conform to the requirement that
 $z_{int}-z_{1}$ contains an integer number of unit cells 
 of the left material,
 and $z_{2}-z_{int}$ contains an integer number of unit cells of 
 the right material. Consequently, the integral of the reference charge density
 $\overline{\rho}_{0}$ between $z_{1}$ and $z_{2}$ carries a net charge.
 On the other hand, as it should, 
 $\overline{\overline{Q}}_{u}$ vanishes and is independent of the
 integration limits provided that $z_{1}$ lies in $R^{'}_{1}$ and
 $z_{2}$ lies in $R^{'}_{2}$.

 \begin{table}
    \caption[ ]{ Interfacial charges and dipole moments of the
                 microscopic interface-like charge density
                 shown in Fig. \ref{fig:nanotoy}(a).
                 The interface charge $Q_{u}$ and the dipole-moment $p_{u}$
                 densities
                 are defined respectively in Eqs. (\ref{eq:qu}) and
                 (\ref{eq:pu}) and the integration limits
                 $z_{1}$ and $z_{2}$ taken midway between the atoms at the 
                 center of each material layer.
                 When a reference charge density 
                 of an ideal interface
                 is subtracted, the corresponding charge 
                 $\Delta Q_{u}$ and dipole moment
                 $\Delta p_{u}$ are defined as in Eqs. (\ref{eq:deltaq}) 
                 and (\ref{eq:deltap}) with the same integration limits as
                 before.
                 Since the nanosmoothed charge density does not play any role
                 in the definition of
                 $Q_{u}$, $p_{u}$, $\Delta Q_{u}$, and $\Delta p_{u}$,
                 these magnitudes are insensitive to nanosmoothing.
                 $\overline{\overline{Q}}_{u}$  and 
                 $\overline{\overline{p}}_{u}$  are defined
		 in Eqs. (\ref{eq:bbqu}) and (\ref{eq:bbpu}).
                 $l_{1}$ and $l_{2}$ are the lengths of the square-wave
                 functions entering in the definition of the
                 smoothing function, Eq. (\ref{eq:filterconvolution}).
               }
    \begin{center}
       \begin{tabular}{cccccccc}
          \hline
          \hline
           $l_{1}$                         &
           $l_{2}$                         &
           $Q_{u}$                         &
           $\Delta Q_{u}$                  &
           $\overline{\overline{Q}}_{u}$   &
           $p_{u}$                         &
           $\Delta p_{u}$                  &
           $\overline{\overline{p}}_{u}$   \\
	   \hline
          $a^{(1)}$ = 6                    &
          $a^{(2)}$ = 8                    &
	  0                                &
	  -0.113                           &
          0                                &
	  1.749                            &
	  -4.821                           &
          2.157                            \\
          $2a^{(1)}$ = 12                   &
          $2a^{(2)}$ = 16                   &
	                                   &
	                                   &
          0                                &
	                                   &
	                                   &
          2.157                            \\
          $3a^{(1)}$ = 18                  &
          $3a^{(2)}$ = 24                  &
	                                   &
	                                   &
          0                                &
	                                   &
	                                   &
          2.157                            \\
          $4a^{(1)}$ = 24                  &
          $4a^{(2)}$ = 32                  &
	                                   &
	                                   &
          0                                &
	                                   &
	                                   &
          2.157                            \\
          $5a^{(1)}$ = 32                  &
          $5a^{(2)}$ = 40                  &
	                                   &
	                                   &
          0                                &
	                                   &
	                                   &
          2.121                            \\
          $\frac{a^{(1)}}{2}$ = 3          &
          $\frac{a^{(2)}}{2}$ = 4          &
	                                   &
	                                   &
          0                                &
	                                   &
	                                   &
          1.871                            \\
          \hline
          \hline
       \end{tabular}
    \end{center}
    \label{table:results-unpol}
 \end{table}

 We have also calculated the corresponding interface dipole densities
 $p_{u}$, Eq. (\ref{eq:pu}); $\Delta p_{u}$, Eq. (\ref{eq:deltap});
 and $\overline{\overline{p}}_{u}$, Eq. (\ref{eq:bbpu}),
 with the results displayed in Table \ref{table:results-unpol}.
 By definition, $p_{u}$ and $\Delta p_{u}$ are independent of the
 nanosmoothing procedure. 
 As long as $l_{1}$ and $l_{2}$ equal an integer number of lattice constants
 of the bulk unit cell of the material along $z$, 
 $\overline{\overline{p}}_{u}$ is insensitive to the shape 
 and range of the smoothing function.
 Note that, as expected after the discussion in
 Sec. \ref{sec:looseinvariance}, the nanosmoothed 
 dipole density $\overline{\overline{p}}_{u}$
 differs from both $\Delta p_{u}$, computed from 
 $\Delta \overline{\rho}_{u} \left( z \right)$,
 and $p_{u}$ computed from $\overline{\rho}_{u} \left( z \right)$. 
 The latter two also differs 
 because $\overline{\rho}_{0}\left( z \right)$ defined in 
 Eq. (\ref{eq:rho0})
 generates an extra dipole-moment density when integrated within 
 our integration limits.
 Thus the electrostatic potential shifts $\Delta \overline{\overline{V}}_{u}$ 
 and $\Delta \overline{V}_{u}$ differ correspondingly.
 The shift $\Delta \overline{\overline{V}}_{u}$ is the physically meaningful one
 because what enters in the band offsets are the local averages of the 
 electrostatic potentials in each material, which are independent
 of the location of $z_{1}$ and $z_{2}$ within 
 $R^{'}_{1}$ and $R^{'}_{2}$ entering in Eq. (\ref{eq:diffdeltaV}) 
 defining $\Delta \overline{\overline{V}}_{u}$.
 The analogous relation $\Delta \overline{V}_{u}$, 

 \begin{equation}
    \Delta \overline{V}_{u} = 
    \overline{V}_{u} \left( z_{2} \right) - 
    \overline{V}_{u} \left( z_{1} \right),  
    \label{eq:deltabV}
 \end{equation}

 \psfrag{  Vbar}[cc][cc]{$\overline{V}_{u} \left( z \right)$}
 \psfrag{Vbarbar}[cc][cc]{$\overline{\overline{V}}_{u} \left( z \right)$}
 \psfrag{zeta1}[cc][cc]{$z_{1}$}
 \psfrag{zeta2}[cc][cc]{$z_{2}$}
 \psfrag{DeltabV}[cc][cc]{$\Delta \overline{V}_{u}$}
 \psfrag{DeltabbV}[cc][cc]{$\Delta \overline{\overline{V}}_{u}$}
 \begin{figure}[htbp]
    \begin{center}
       \includegraphics[width=10cm] {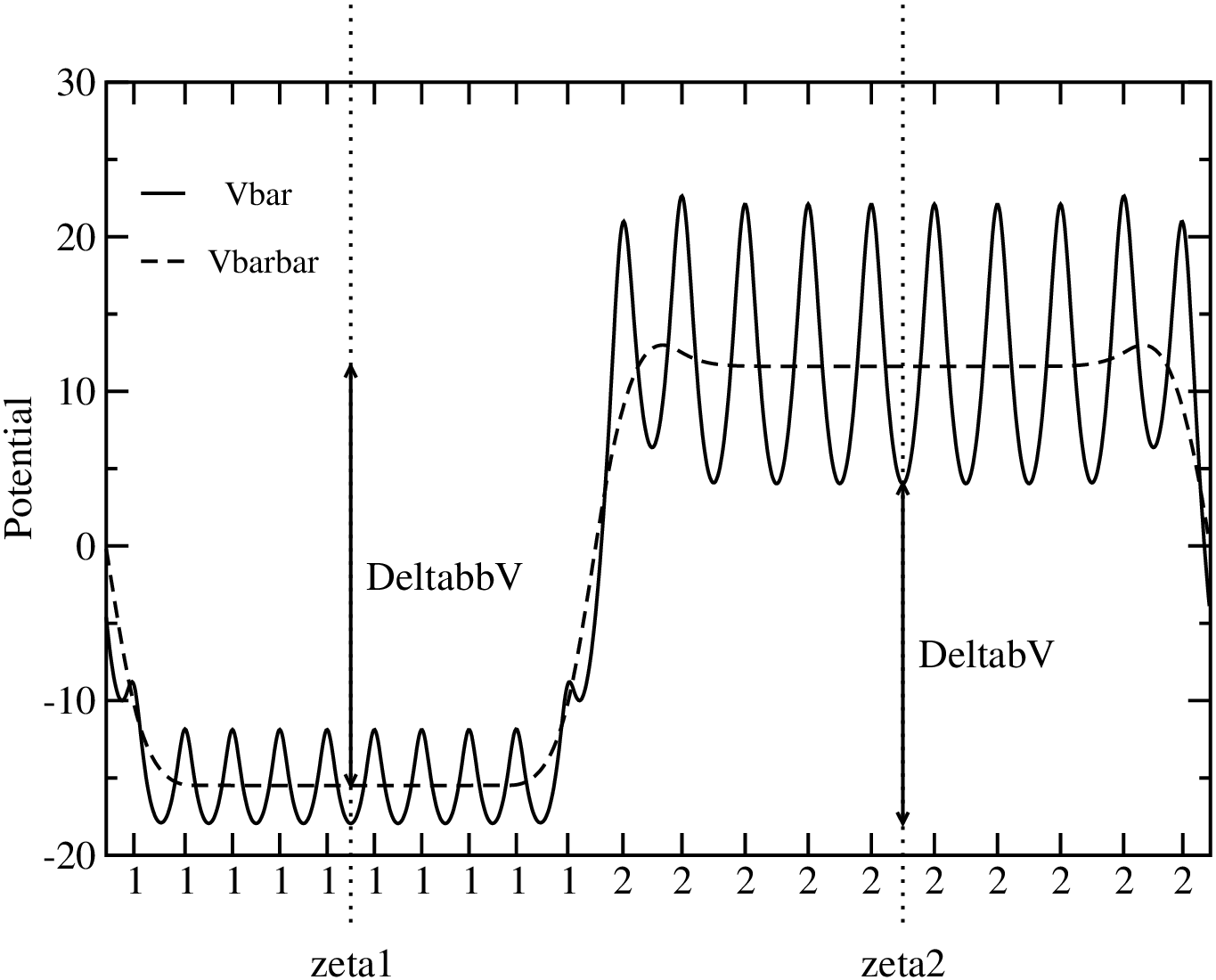}
       \caption{ Planar averaged electrostatic potential computed by solving 
                 the one dimensional Poisson equation,  
                 Eq. (\ref{eq:transavpoisson}),
                 with the charge density 
                 $\overline{\rho}_{u} \left( z \right)$ shown in 
                 Fig. \ref{fig:nanotoy}(a), 
                 $\overline{V}_{u} \left( z \right)$, 
                 and by solving Eq. (\ref{eq:poissonnanosmoothed}) 
		 with the charge density 
                 $\overline{\overline{\rho}}_{u} \left( z \right)$ shown in
                 Fig. \ref{fig:nanotoy}(c),
                 $\overline{\overline{V}}_{u} \left( z \right)$.
                 The integration limits $z_{1}$ in $R^{'}_{1}$ 
                 and $z_{2}$ in $R^{'}_{2}$ are indicated by dotted lines.
                 The electrostatic potential shifts 
                 $\Delta \overline{V}_{u}$ [Eq. (\ref{eq:deltabV})]  
                 and $\Delta \overline{\overline{V}}_{u}$ 
                 [Eq. (\ref{eq:diffdeltaV})]
                 defined as the difference of
                 the corresponding potentials at the points $z_{1}$ and
                 $z_{2}$ are also shown.
                 Atomic units are used. 
         }
       \label{fig:potential}
    \end{center}
 \end{figure}

 \noindent shows that $\Delta \overline{V}$ is the difference of the
 two potentials at specific points within regions
 in which $\overline{V} \left( z \right)$ varies rapidly.
 In Fig. \ref{fig:potential} we show $\overline{\overline{V}} \left( z \right)$
 and $\overline{V} \left( z \right)$ and indicate
 the positions $z_{1}$ and $z_{2}$.
 It is clear that $\Delta \overline{\overline{V}}$ is 
 about 20\% larger than $\Delta \overline{V}$, explaining the
 relation between $\overline{\overline{p}}_{u}$ and $p_{u}$
 in Table \ref{table:results-unpol},
 the difference arising from the way $\overline{V}$ is sampled
 by nanosmoothing and by the selection of $z_{1}$ and $z_{2}$.

 In Table \ref{table:results-unpol} we also show two cases where the 
 nanosmoothing does affect the value $\overline{\overline{p}}_{u}$.
 First, the value of $\overline{\overline{p}}_{u}$
 departs from the correct value 2.157
 when the range of the smoothing function, $L$ 
 in Eqs. (\ref{eq:filterfuna}) and (\ref{eq:filterfunb}), is of the same 
 magnitude as the width of the layer of one of the materials,
 as for $l_{1} = 5a^{(1)} = 30$ and $l_{2} = 5a^{(2)} = 40$,
 so $L = l_{1} + l_{2} = 70$, slightly larger than the width of material 1,
 made of 10 layers with an interlayer distance $a^{(1)} = 6$.
 Under these circumstances, it is not possible to define regions
 $R^{'}_{1}$ and $R^{'}_{2}$ where 
 $\overline{\overline{\rho}}_{u} \left( z \right)$ vanishes, 
 impeding a proper location of 
 the integration limits $z_{1}$ and $z_{2}$. 
 Second, the same occurs 
 when the range of every filter function entering
 Eq. (\ref{eq:filterconvolution}) does not equal an integer
 number of lattice constants of the bulk unit cell along $z$,
 as for $l_{1} = \frac{a^{(1)}}{2} = 3$, and $l_{2} = \frac{a^{(2)}}{2} = 4$.
 In such a case, the charge density after nanosmoothing, 
 $\overline{\overline{\rho}}_{u} \left( z \right)$, still shows large
 and rapid oscillations. 

\begin{figure}[htbp]
    \begin{center}
       \includegraphics[width=10cm] {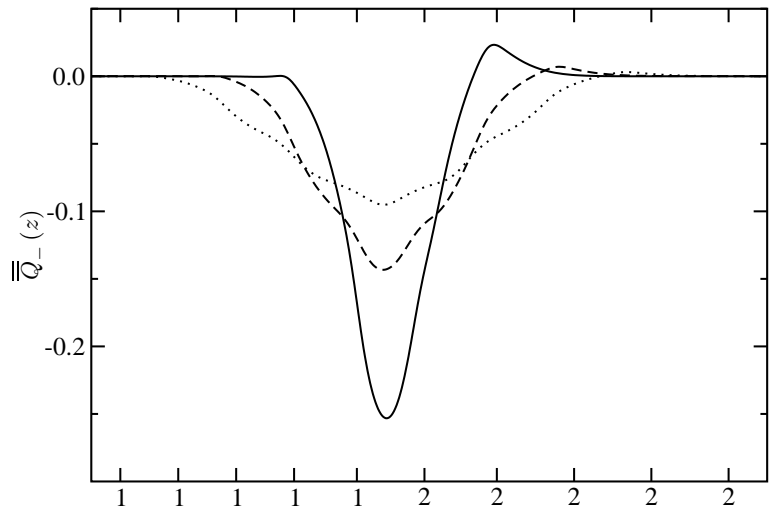}
       \caption{ Negative cumulative charges defined in Eq. (\ref{eq:cumul-})
                 for the nanosmoothed charge densities shown in 
                 Fig. \ref{fig:diffr}.
                 We can estimate the magnitude of the charge transferred
		 per unit area of the interface $q$ from the supremum of 
		 this cumulative charge and the interface position from 
		 the position where this supremum appears.
		 Numerical values are given in Table \ref{table:posint}.
         }
       \label{fig:posint}
    \end{center}
 \end{figure}

 In Table \ref{table:posint} we report the amount of charge transferred
 from one material to the other, the screening length and the interface
 position computed with the method summarized in 
 Sec. \ref{sec:nanosmoothing-transferch} 
 for the nanosmoothed charge densities plotted in Fig. \ref{fig:diffr}.
 The negative cumulative charges are displayed in Fig. \ref{fig:posint}.
 As expected the value of $q$ is sensitive to the smoothing function,
 decreasing with the increasing of its range. 
 Accordingly, the screening length is also sensitive to the 
 range of the smoothing function $L$, taking a value that is roughly 
 the greater of $l_{1}$ and $l_{2}$.
 Nevertheless, the position of the interface seems to be insensitive
 to the filtering function.

 The results displayed in Fig. \ref{fig:diffr} for
 $\overline{\overline{\rho}}(z)$ demonstrate that the larger 
 the width of the smoothing function, the more complex the
 spatial dependence of $\overline{\overline{\rho}}(z)$
 and the less it resembles a simple interface charge density. 
 The introduction of multiple extrema is artificial and caused
 by excessive transfer of charge back across the interface 
 by smoothing. This back transfer of charge is manifested clearly in the
 values of $q$ in Table \ref{table:posint}.
 Accordingly {\it the smallest allowable widths $l_{s} = a_{s}$,
 should be used for nanosmoothing.} The resulting values 
 of $q$ and $\lambda$ are the best estimates which can be extracted by 
 nanosmoothing.

 \psfrag{Cumu}[cc][cc]{$\overline{\overline{Q}}_{-} \left( z \right)$}

 \begin{table}
    \caption[ ]{ Magnitude of the transferred charge density, 
                 screening length and
                 position of the interface computed from the 
                 nanosmoothed charge densities plotted in Fig. \ref{fig:diffr}.
                 $l_{1}$ and $l_{2}$ are the lengths of the square-wave
                 functions entering in the definition of the
                 smoothing function, Eq. (\ref{eq:filterconvolution}).
                 Since the interface is unpolarized, the charge at each
                 side of the interface must be equal in magnitude and oposite
                 in sign.
               }
    \begin{center}
       \begin{tabular}{ccccc}
          \hline
          \hline
          $l_{1}$       &
          $l_{2}$       &
          $z_{int}$     &
          $q$           &
          $\lambda$     \\
          \hline
          6             &
          8             &
          58.083        &
          0.253         &
          8.517         \\
          12            &
          16            &
          57.709        &
          0.143         &
          15.041        \\
          18            &
          24            &
          57.771        &
          0.095         &
          22.660        \\
          \hline
          \hline
       \end{tabular}
    \end{center}
    \label{table:posint}
 \end{table}

 Now, as a final test of the insensitivity of the dipole-moment density
 and potential shift to the shape of the smoothing function, we shall simplify
 the toy model to one with 
 a common atomic separation in both materials.
 This opens three simple
 options for the filter function.
 We can set $f$ equal to a single $\omega$ when the filter function
 adopts a square shape [see Fig. \ref{fig:filter}(a)], we can set
 $f$ equal to the convolution of two $\omega$'s with the same $l$,
 a triangle function, 

 \begin{equation}
   f ( z - z^{'} ) = \frac{1}{l^{2}} \left( l - \left|z - z^{'} \right| \right),
 \end{equation}

 \noindent or we can set $f$ equal to the 
 convolution of two $\omega$'s with different $l$,
 a trapezoidal function, as in Eq. (\ref{eq:doublefilter}) 
 and in Fig. \ref{fig:filter}(b).
 Comparing the results of the smoothing with the square wave, the triangle, and
 the trapezoidal function for various values of $l$ allows us to 
 establish which physical properties are insensitive both to the range
 and to the shape of the smoothing function.

 \begin{table}
    \caption[ ]{ Interfacial charge and dipole moment densities for a 
                 microscopic interface-like charge density obtained
                 with the following parameters:
                 $A^{n,1} = A^{e,1} = 1$, 
                 $\sigma_{n,1} = 0.5$,
                 $\sigma_{e,1} = 2.0$,
                 $A^{n,2} = A^{e,2} = 2$, 
                 $\sigma_{n,2} = 0.7$,
                 $\sigma_{e,2} = 4.5$, 
                 $\delta^{(1)} = \delta^{(2)} = 0$ for all the ``atoms'', 
	         $a^{(1)} = 10$, and  
	         $a^{(2)} = 10$.
                 The meaning of the symbols and integration limits is
		 as in Table \ref{table:results-unpol}.
               }
    \begin{center}
       \begin{tabular}{lcccccccc}
          \hline
          \hline
                                           &
           $l_{1}$                         &
           $l_{2}$                         &
           $Q_{u}$                         &
           $\Delta Q_{u}$                  &
           $\overline{\overline{Q}}_{u}$   &
           $p_{u}$                         &
           $\Delta p_{u}$                  &
           $\overline{\overline{p}}_{u}$   \\
           \hline
 	  Square filter function          &
 	  10    &
 	  -     &
 	  0     &
 	  0     &
 	  0     &
 	  1.202 &
 	  1.202 &
 	  1.813 \\
 	        &
 	  50    &
 	  -     &
 	        &
 	        &
 	  0     &
 	        &
 	        &
 	  1.813 \\
  Triangle filter function        &
          10    &
          10    &
          0     &
          0     &
          0     &
          1.202 &
          1.202 &
          1.813 \\
                &
          20    &
          20    &
                &
                &
          0     &
                &
                &
          1.813 \\
  Trapezoidal filter function    &
          10    &
          20    &
          0     &
          0     &
          0     &
          1.202 &
          1.202 &
          1.813 \\
                &
          5     &
          10    &
                &
                &
          0     &
                &
                &
          1.813 \\
          \hline
          \hline
       \end{tabular}
    \end{center}
    \label{table:results-unpol-filter}
 \end{table}

 Again, as soon as $l_{1}$ and/or $l_{2}$ equal an 
 integer number of lattice constants
 of the bulk unit cell of the relevant material along $z$, 
 $\overline{\overline{p}}_{u}$ is insensitive to the shape (triangular, square
 or trapezoidal)
 and range of the smoothing function.

 \section{Polarized systems.}
 \label{sec:polar}

 Up to now we have dealt with materials unpolarized except at the interfaces.
 In such systems it was possible to identify bulk-like regions in the middle
 of each layer where the microscopic charge density is unaffected
 by the presence of interfaces and returns to the bulk microscopic
 charge density, which is centrosymmetric within a unit cell.
 We now turn our attention to polarized interfaces, where at 
 least one of the materials has a non-vanishing polarization
 $\vec{P}$.

 Basic electrostatic arguments that can be found in any textbook
 \cite{Feynman,Jackson} show that a non uniform polarization in a dielectric
 generates a volume charge density, the polarization charge
 $\rho_{pol} \left( \vec{r} \right)$, whose value
 at any point of space is given by

 \begin{equation}
    \rho_{pol} \left( \vec{r} \right) = 
    - \nabla \cdot \vec{P} \left( \vec{r} \right).
 \end{equation}

 \noindent Even in the case of uniformly polarized material,
 where $\vec{P}$ is constant and therefore its divergence
 vanishes inside the dielectric, the discontinuity of the polarization
 at the surface or interface gives rise to a net surface or interface
 charge density $\sigma_{pol}$ given by the familiar form

 \begin{equation}
    \sigma_{pol} = \vec{P} \cdot \hat{n},
 \end{equation}

 \noindent where $\hat{n}$ is a unit vector normal to the surface or
 interface pointing outwards.

 In the case of an interface, 
 the other material at the interface responds to this perturbation
 in order to minimize the electrostatic energy cost associated
 with the build up of the polarization charge $\nabla \cdot \vec{P}$
 at the interfaces.
 If the second material is a dielectric, a uniform polarization is induced
 within it as well.  \cite{Neaton-03, Nakhmanson-05}
 If the second material is a metal, a screening charge is induced
 that spreads over a finite distance (the screening length) in the electrode,
 producing additional dipole layers at each polar dielectric/metal interface.
 If the screening of the polarization charge is not perfect,
 a residual depolarizing field appears inside the 
 dielectric.\cite{Ghosez-06,Junquera-03.1}

 We now face the same question as for the unpolarized case:
 how to extract from the results of the first-principles calculations
 values of the physical quantities of interest (polarization and
 screening charge densities, screening lengths, depolarizing fields, etc). 

 In parallel to what was done for the unpolarized case,
 we can define the net charge density $Q_{p}$ and 
 dipole moment density $p_{p}$ associated with the 
 microscopic charge density of the polar interface $\overline{\rho}_p$ as

 \begin{subequations}
    \begin{align}
      Q_p & = \int_{z_{1}}^{z_{2}} dz \:\: \overline{\rho}_p \left( z \right), 
      \label{eq:qp} 
      \\
      p_p & = \int_{z_{1}}^{z_{2}} dz \:\: z \overline{\rho}_p \left( z \right).
      \label{eq:pp}
    \end{align}
 \end{subequations}

 \noindent The computation of these two quantities present the 
 same difficulties as for the unpolarized case, 
 summarized in Sec. \ref{sec:unpolar}.

 The approach of determining a reference interface charge density
 $\overline{\rho}_{0} \left( z \right)$ as in Eq. (\ref{eq:rho0})
 and defining an interface-induced deformation of the charge density
 $\Delta \overline{\rho}_{p} \left( z \right)$ by subtracting 
 $\overline{\rho}_{0} \left( z \right)$ from 
 the microscopic charge density, 

 \begin{equation}
    \Delta \overline{\rho}_{p} \left( z \right) = 
    \overline{\rho}_{p} \left( z \right) - \overline{\rho}_{0} \left( z \right),
    \label{eq:deltarhop}
 \end{equation}

 \noindent discussed in Sec. \ref{sec:refint} for the unpolarized case, 
 still allows for the
 identification of two regions $R_{1}$ and $R_{2}$ where
 $\Delta \overline{\rho}_{p} \left( z \right)$
 vanishes [see Fig. \ref{fig:nanotoypolar}(b)]. 
 Therefore we can define both the interface charge and dipolar
 densities invariant with respect the position of the integration
 limits $z_{1}$ and $z_{2}$, as long as $z_{1}$ lies in $R_{1}$ and
 $z_{2}$ lies in $R_{2}$.

 However, on top of all the drawbacks to that 
 approach presented in Sec. \ref{sec:refint},
 there is an additional pitfall now:
 an extra dependence on the position of the 
 integration limits $z_{1}$ and $z_{2}$ appears in the computation 
 of the interface dipole moment from the definition of the 
 reference interface position $z_{int}$. 
 Indeed, for at least one of the materials the dipole moment density does not
 vanish within the bulk unit cell, 

 \begin{equation}
    p^{(s)}_{0} = \int_{a^{(s)}_{bulk}} dz \:\: 
    z \overline{\rho}^{(s)}_{0} \left( z \right).
 \end{equation}

 \noindent Therefore, the dipole moment density $\Delta p_{p}$ associated with 
 $\Delta \overline{\rho}_{p} \left( z \right)$ does not equal $p_{p}$ defined
 in Eq. (\ref{eq:pp}), not even in the case where 
 $z_{int} - z_{1}$ contains an integer number $M_{1}$ of unit cells of the 
 left material and $z_{2}-z_{int}$ contains an integer number $M_{2}$
 of unit cells of the right material,

 \begin{align}
    \Delta p_{p} & = \int_{z_{1}}^{z_{2}} dz \:\: 
                   z \Delta \overline{\rho}_p \left( z \right) =
		     \int_{z_{1}}^{z_{2}} dz \:\:
                   z \left[ \overline{\rho}_{p} \left( z \right) - 
                          \overline{\rho}_{0} \left( z \right)
		   \right]
   \nonumber \\
        & = \int_{z_{1}}^{z_{2}} dz \:\: z \overline{\rho}_{p} \left( z \right)
          - \int_{z_{1}}^{z_{2}} dz \:\: z \overline{\rho}_{0} \left( z \right)
   \nonumber \\
        & = p_{p} - M_{1} p^{(1)}_{0} - M_{2} p^{(2)}_{0}.
   \label{eq:deltapp}
 \end{align}

 Nevertheless, the nanosmoothing procedure developed in 
 Sec. \ref{sec:nanosmoothing-procedure}
 remains a useful tool, since 
 the methodology for constructing the nanosmoothed charge density,
 Eq. (\ref{eq:nanosmoothing}), is independent on whether 
 the microscopic charge density is polarized or not.
 Both the nanosmoothed Poisson equation [Eq. (\ref{eq:poissonnanosmoothed})] 
 and the analysis of the insensitivity of the dipole
 moment density to the smoothing function 
 (Sec. \ref{sec:nanosmoothing-insensitivity}) still hold under
 the same conditions for the smoothing function as for the unpolarized case, 
 because the fundamental starting point for the derivation,
 the microscopic Poisson equation and the existence of regions with 
 a negligible nanosmoothed charge $R^{'}_{1}$ and $R^{'}_{2}$, 
 are insensitive to the polarization state of the microscopic charge density.

 \section{Results: toymodel, polarized.}
 \label{sec:resultspolar}

\begin{figure}[htbp]
    \begin{center}
       \includegraphics[width=10cm]{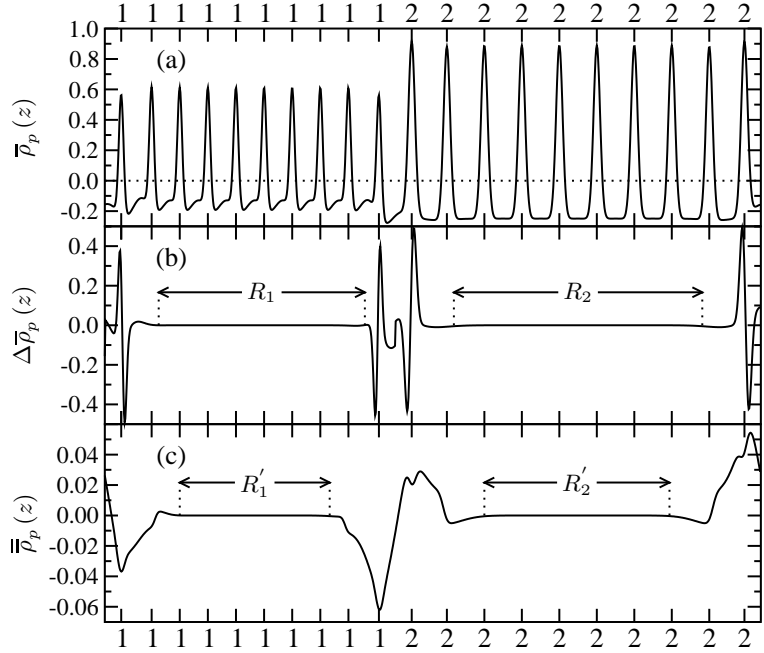}
       \caption{ (a) Microscopic charge density 
                 $\overline{\rho}_{p} \left( z \right)$ within a unit supercell
                 for a polarized system, with the polarization induced
                 by a displacement of the electronic charge in one of the
                 materials.
                 The widths of the layers of both material and the rest of
                 the parameters of the toy model are as in the caption of Fig.
                 \ref{fig:nanofun}, with the exception of 
                 $\delta^{(1)}_{i} = 1.0$, where $i$ runs over all the ``atoms''
                 of material 1. 
                 (b) Interface-induced deformation of the charge density 
                 $\Delta \overline{\rho}_{p} \left( z \right)$ 
                 defined as in Eq. (\ref{eq:deltarhop}).
                 As before, the interface plane $z_{int}$ is positioned at
                 the middle of the
                 separation between the last atomic plane on the
                 left and the first atomic plane on the right.
                 (c) Nanosmoothed charge density 
                 $\overline{\overline{\rho}}_{p}(z)$ of the microscopic charge 
                 density. The nanosmoothing function is defined as in 
                 Eq. (\ref{eq:filterconvolution}) with 
                 $l_{1} = a^{(1)} = 6$, and 
                 $l_{2} = a^{(2)} = 8$,  single interplanar 
                 distances along $z$ for each material.
		 Atomic units are used.
         }
       \label{fig:nanotoypolar}
    \end{center}
 \end{figure}

 In Fig. \ref{fig:nanotoypolar}(a) we illustrate the first specific 
 toy model we shall analyze in detail for the polarized case.
 The parameters of the model are the same as in the caption of 
 Fig. \ref{fig:nanofun} with the exception of the parameter
 $\delta^{(1)}_{i}$, that now takes a non-zero value 
 $\delta^{(1)}_{i} = 1.0$ where $i$ runs over all the ``atoms''
 of material 1. This toy model simulates a multilayer in which 
 a net polarization has been induced in material 1
 by displacing the ``electronic clouds'' rigidly 1.0 length units towards the
 right. 
 The neutral but polarized atom of this toy model can be thought of as
 representing the overall neutral atomic planes of a ferroelectric
 within which a dipole-moment density is generated 
 by buckling or puckering.
 In principle, the charge density of material 2 would be modified as a response
 to the presence of a polarization in material 1. This polarization-induced 
 response, which should be computed self-consistently,
 is not considered in the present simple toy model. 
 The displacement of the negative charges translates into the asymmetry of the 
 charge density inside material 1 represented in Fig. \ref{fig:nanotoypolar}(a).
 The interface-induced deformation of the charge density 
 $\Delta \overline{\rho}_{p} \left( z \right)$, defined as in 
 Eq. (\ref{eq:deltarhop}), 
 is shown in Fig. \ref{fig:nanotoypolar}(b) where, as for the unpolarized case,
 the position of the interface $z_{int}$ has been located half way between
 the rightmost left atomic layer and the leftmost right atomic layer.
 The nanosmoothed charge density 
 $\overline{\overline{\rho}}_{p} \left( z \right)$ is displayed in
 Fig. \ref{fig:nanotoypolar}(c), obtained with the same smoothing
 function as the the one used in the construction of
 $\overline{\overline{\rho}}_{u} \left( z \right)$ 
 in Fig. \ref{fig:nanotoy}(c).
 In the last two panels, the two regions inside materials 1 and 2
 where $\Delta \overline{\rho}_{p} \left( z \right)$ and
 $\overline{\overline{\rho}}_{p} \left( z \right)$ vanish
 are clearly shown.

 \psfrag{Yaxisp1}[cc][cc]{$\overline{\rho}_{p} \left( z \right)$}
 \psfrag{Yaxisp2}[cc][cc]{$\Delta \overline{\rho}_{p} \left( z \right)$}
 \psfrag{Yaxisp3}[cc][cc]{$\overline{\overline{\rho}}_{p} \left( z \right)$}

 Some of the features already discussed in Sec. \ref{sec:resultsunpolar}
 for the unpolarized interface still remain valid for the polarized case.
 In particular: $(i)$
 the regions $R^{'}_{s}$ within which $\overline{\overline{\rho}}_{p}$
 vanishes lie within $R_{s}$ within which $\Delta \overline{\rho}_{p}$
 vanishes, because smoothing $\overline{\rho}_{p}$ 
 within $R_{s}$ brings into $\overline{\overline{\rho}}_{p}$
 values of $\overline{\rho}_{p}$ for $z$ outside $R_{s}$;
 $(ii)$ the discontinuity of $\Delta \overline{\rho}_{p}$ 
 at the interface and the large fluctuations at the atomic scale in the
 neighborhood of the interface;
 $(iii)$ the smooth and continuous profile [note the change of scale in the
 charge density in Fig. \ref{fig:nanotoypolar}(b) and 
 Fig. \ref{fig:nanotoypolar}(c)] of
 $\overline{\overline{\rho}}_{p}$, highlighting its advantages
 with respect $\Delta \overline{\rho}_{p}$.
 However, some other issues are new. Now, 
 as a consequence of the asymmetry of
 $\overline{\rho}_{p} \left( z \right)$,
 the charge density at the interface regions, defined as the ranges 
 where 
 $\Delta \overline{\rho}_{p} \left( z \right)$ and
 $\overline{\overline{\rho}}_{p} \left( z \right)$ 
 differ significantly from zero,
 are different in the adjacent interfaces contained in our simulated supercell
 [see the center and the edge of Fig. \ref{fig:nanotoypolar}(b) 
 and \ref{fig:nanotoypolar}(c)].
 They are not simply reflections of one another as in 
 Fig. \ref{fig:nanotoy}(b) and \ref{fig:nanotoy}(c).

 In Table \ref{table:results-pol} we report the values
 of the interface charges 
 computed from 
 $\overline{\rho}_{p} \left( z \right)$ ($Q_{p}$),
 $\Delta \overline{\rho}_{p}\left( z \right)$ 
 ($\Delta Q_{p}$),
 and $\overline{\overline{\rho}}_{p} \left( z \right)$ 
 ($\overline{\overline{Q}}_{p}$). 
 The integration limits $z_{1}$ and $z_{2}$ are taken midway between 
 the atoms at the center of each material layer.
 In contrast to the unpolarized interface, $Q_{p}$
 does not vanish. 
 The surplus of negative charge is due to the charge density 
 entering region $z_{2}-z_{int}$ from the left from
 the region between $z_{int}-z_{1}$ which is
 not compensated by the departure of any charge. 
 Since $\overline{\rho}_{p} \left( z \right)$ is not even 
 about $z_{1}$, $\overline{\overline{Q}}_{p} \ne Q_{p}$.
 As happened in the unpolarized case,
 as long as $l_{1}$ and $l_{2}$ equal an integer number of lattice constants
 of the bulk unit cell of the material along $z$, $\overline{\overline{Q}}_{p}$
 and $\overline{\overline{p}}_{p}$ are insensitive to the shape and range
 of the smoothing function.
 Only when the range of every filter function entering in 
 Eq. (\ref{eq:filterconvolution}) does not equal an integer number of lattice
 constants of the bulk unit cell, or when the range of the smoothing function
 $L$ is of the same order as the width of the layer of one of the materials,
 do $\overline{\overline{Q}}_{p}$ and $\overline{\overline{p}}_{p}$
 differ from the correct value.

\begin{figure}[htbp]
    \begin{center}
       \includegraphics[width=10cm]{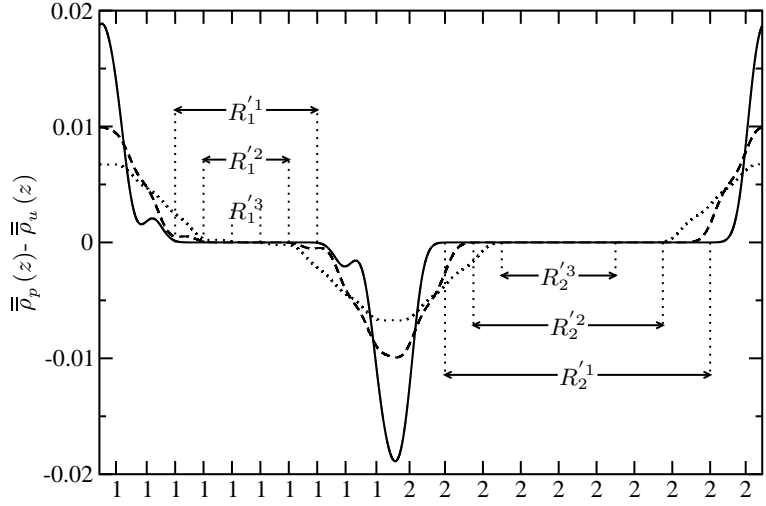}
       \caption{ Difference between the nanosmoothed charge density for the
                 polarized interface and an unpolarized interface.
                 The parameters of the microscopic charge density as 
                 are in Figs. \ref{fig:nanotoypolar} and \ref{fig:nanofun},
                 respectively.
                 Three different filter functions are used for the 
                 nanosmoothing, with the same parameters as in 
                 Fig. \ref{fig:diffr}.
         }
       \label{fig:sigmapol}
    \end{center}
 \end{figure}

 Subtracting from  $\overline{\overline{\rho}}_{p} \left( z \right)$ 
 the related $\overline{\overline{\rho}}_{u} \left( z \right)$,
 a microscopic charge density 
 constructed by nanosmoothing with filter functions
 defined with the same parameters
 with the exception of the displacement of the electronic charge,
 we obtain the profile of the nanosmoothed polarization-induced 
 charge density, Fig. \ref{fig:sigmapol}, 
 whose integral between $z_{1}$ and $z_{2}$ gives $\sigma_{pol}$ as the
 accumulation of charge at the interface.
 Note that since the resulting $\overline{\overline{Q}}_{u}$ vanishes,
 $\overline{\overline{Q}}_{p}$ is identical to $\sigma_{pol}$.
 Although, as listed in Table \ref{table:results-pol},
 the amount of charge accumulated at the interface is independent
 of the shape and range of the filter function used
 (provided that $l_{1}$ and $l_{2}$ are integer 
 numbers of lattice constants of the bulk unit cell of the material along $z$),
 the profile of the polarization-induced charge density is not.
 The wider the nanosmoothing function, the larger the range of the
 polarization charge density. 
 Again, this sensitivity of the shape of the smoothed charge density
 to the nanosmoothing function spoils a direct physical 
 interpretation of its features. 
 The smallest values of $l_{1}$ and $l_{2}$, a single lattice 
 constant of each material, yields, 
 as for the unpolarized case, the simplest and physically most
 relevant profile for $\overline{\overline{\rho}}_{p}$.
 
 \psfrag{Yaxisp4}[cc][cc]{$\overline{\overline{\rho}}_{p} \left( z \right)$-
 $\overline{\overline{\rho}}_{u} \left( z \right)$}

\begin{figure}[htbp]
    \begin{center}
       \includegraphics[width=10cm]{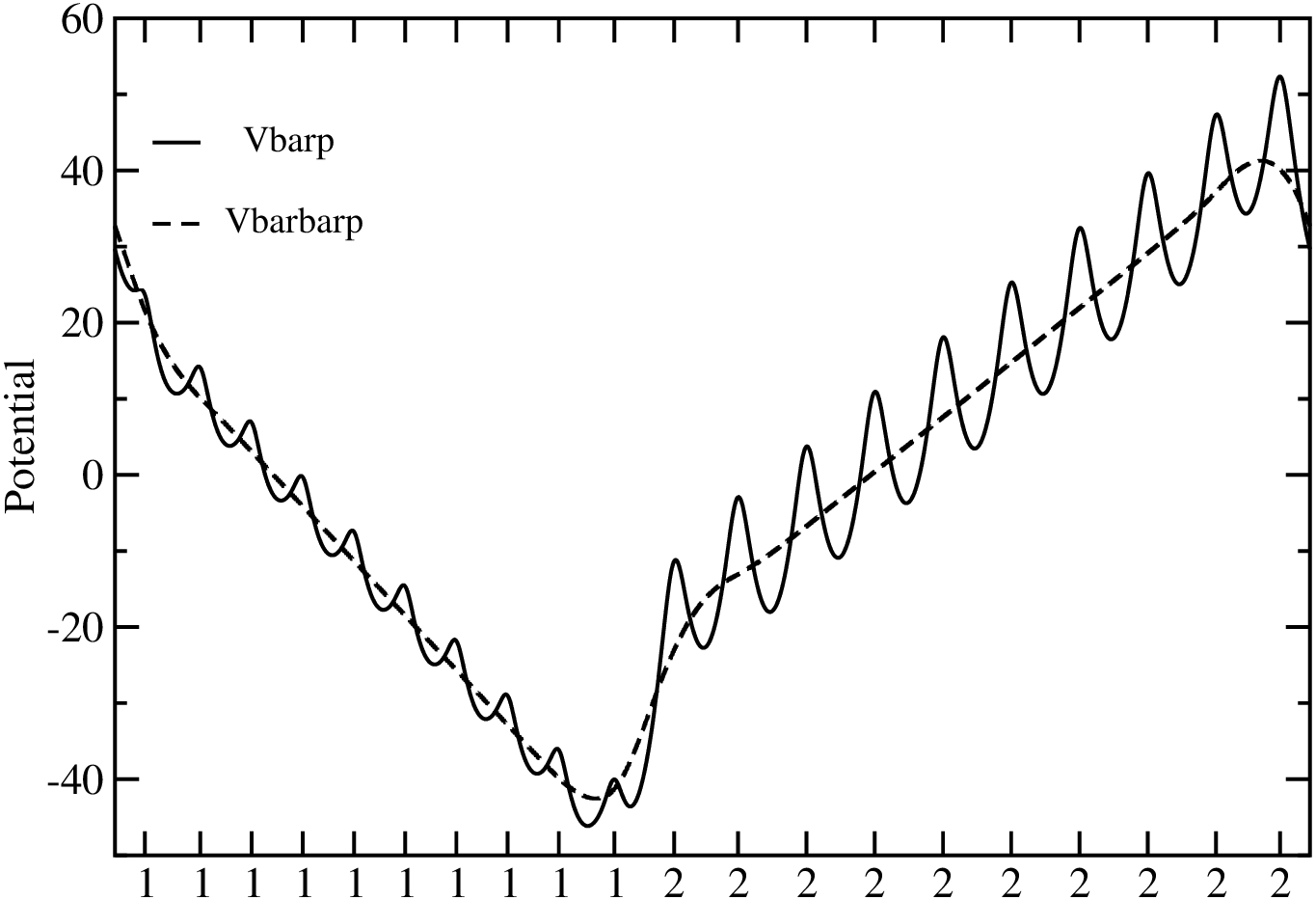}
       \caption{ Planar averaged electrostatic potential computed by solving 
                 the one dimensional Poisson equation,  
                 Eq. (\ref{eq:transavpoisson}),
                 with the charge density 
                 $\overline{\rho}_{p} \left( z \right)$ shown in 
                 Fig. \ref{fig:nanotoypolar}(a), 
                 $\overline{V}_{p} \left( z \right)$, 
                 and by solving Eq. (\ref{eq:poissonnanosmoothed}) 
		 with the charge density 
                 $\overline{\overline{\rho}}_{p} \left( z \right)$ shown in
                 Fig. \ref{fig:nanotoypolar}(c),
                 $\overline{\overline{V}}_{p} \left( z \right)$.
                 The nanosmoothing function is defined as in 
                 Eq. (\ref{eq:filterconvolution}) with 
                 $l_{1} = a^{(1)} = 6$, and 
                 $l_{2} = a^{(2)} = 8$,  single interplanar 
                 distances along $z$ for each material.
                 Atomic units are used.
         }
       \label{fig:polarpotential}
    \end{center}
 \end{figure}

\begin{figure}[htbp]
    \begin{center}
       \includegraphics[width=10cm]{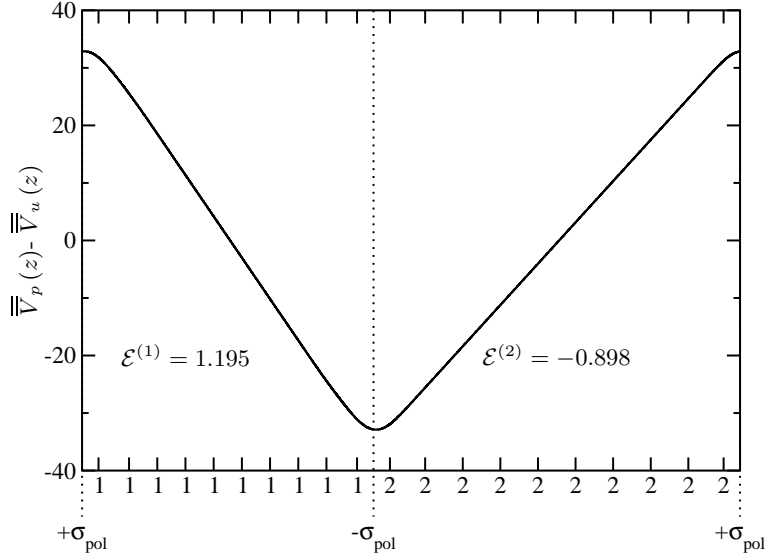}
       \caption{ Difference of the nanosmoothed potential
                for the polarized (dashed line in Fig. \ref{fig:polarpotential})
                 and unpolarized (dashed line in Fig. \ref{fig:potential})
                 cases. The magnitude of the 
		 electric field $\mathcal{E}$ generated by a periodic
                 array of polarization charge distributions $\sigma_{pol}$
                 can be computed from the slope of the profile of the potential
                 within material 1 and material 2.
                 Atomic units are used.
         }
       \label{fig:field}
    \end{center}
 \end{figure}

 Solving the nanosmoothed Poisson equation, Eq. (\ref{eq:poissonnanosmoothed}),
 for the nanosmoothed charge density of Fig. \ref{fig:nanotoypolar}(a)
 we obtain the nanosmoothed potential 
 $\overline{\overline{V}}_{p} \left( z \right)$
 displayed in Fig. \ref{fig:polarpotential}.
 This electrostatic potential has two distinct features:
 $(i)$ a jump at each interface
 due to the interface dipole moment already present 
 in the unpolarized case (see Fig. \ref{fig:potential} 
 and Table \ref{table:results-unpol});
 $(ii)$ a field generated by the polarization charge 
 $\sigma_{pol}$ at the interface.
 We can isolate this field by subtracting the
 nanosmoothed potential for the polarized (Fig. \ref{fig:polarpotential})
 and unpolarized (Fig. \ref{fig:potential}) systems.
 The result is shown in Fig. \ref{fig:field}.
 The electric field can be computed from the slope of the 
 nanosmoothed potential.

 \psfrag{  Vbarp}[cc][cc]{$\overline{V}_{p} \left( z \right)$}
 \psfrag{Vbarbarp}[cc][cc]{$\overline{\overline{V}}_{p} \left( z \right)$}

 \psfrag{Yaxisp5}[cc][cc]{$\overline{\overline{V}}_{p} \left( z \right)$-
 $\overline{\overline{V}}_{u} \left( z \right)$}
 \psfrag{Efield1}[cc][cc]{$\mathcal{E}^{(1)} = 1.195$}
 \psfrag{Efield2}[cc][cc]{$\mathcal{E}^{(2)} = -0.898$}

 The magnitude of the electric field is related to the magnitude of the
 polarization charge $\sigma_{pol}$. 
 The periodic boundary conditions enforces
 that $\mathcal{E}^{(1)} N_{1} a^{(1)}_{bulk} + 
 \mathcal{E}^{(2)} N_{2} a^{(2)}_{bulk} = 0$,
 where $N_{1}$ and $N_{2}$ are the number of unit cells of materials
 1 and 2 stacked to build the superlattice (see Sec. \ref{sec:toymodel}).
 Since each interface carries a charge of magnitude $\sigma_{pol}$,
 the electric field changes its value at the interface as
 $\mathcal{E}^{(2)} - \mathcal{E}^{(1)} = 4 \pi \sigma_{pol}$, yielding
 to  values for the fields
 $\mathcal{E}^{(1)} = -\frac{4 \pi \sigma_{pol} N_{2} a^{(2)}_{bulk}}
 {N_{1}a^{(1)}_{bulk}+N_{2}a^{(2)}_{bulk}}$
 and 
 $\mathcal{E}^{(2)} = \frac{4 \pi \sigma_{pol} N_{1}a^{(1)}_{bulk}}
 {N_{1}a^{(1)}_{bulk}+N_{2}a^{(2)}_{bulk}}$.
 Taking $\sigma_{pol}$ from Table \ref{table:results-pol},
 we infer a value of $\mathcal{E}^{(1)}$ = 1.196, and
 $\mathcal{E}^{(2)} = -0.897$, in excellent agreement
 with the slopes computed in Fig. \ref{fig:field}.

 \begin{table}
    \caption[ ]{ Interfacial charges and dipole moment densities of the 
                 microscopic interface-like charge density for the polar
                 interface shown in Fig. \ref{fig:nanotoypolar}(a).
                 The meaning of the different symbols is as in the caption of
                 Table \ref{table:results-unpol}, with the 
		 subscript $u$ replaced by $p$ 
                 for the polar case.
               }
    \begin{center}
       \begin{tabular}{cccccccc}
          \hline
          \hline
           $l_{1}$                         &
           $l_{2}$                         &
           $Q_{p}$                         &
           $\Delta Q_{p}$                  &
           $\overline{\overline{Q}}_{p}$   &
           $p_{p}$                         &
           $\Delta p_{p}$                  &
           $\overline{\overline{p}}_{p}$   \\
	   \hline
          $a^{(1)}$ = 6                    &
          $a^{(2)}$ = 8                    &
	  -0.136                           &
	  -0.249                           &
          -0.167                           &
	  -7.055                           &
	  -12.708                          &
          -7.509                           \\
          $2a^{(1)}$ = 12                   &
          $2a^{(2)}$ = 16                   &
	                                   &
	                                   &
          -0.167                           &
	                                   &
	                                   &
          -7.509                           \\
          $3a^{(1)}$ = 18                  &
          $3a^{(2)}$ = 24                  &
	                                   &
	                                   &
          -0.167                           &
	                                   &
	                                   &
          -7.509                           \\
          $4a^{(1)}$ = 24                  &
          $4a^{(2)}$ = 32                  &
	                                   &
	                                   &
          -0.166                           &
	                                   &
	                                   &
          -7.754                           \\
          $\frac{a^{(1)}}{2}$ = 3          &
          $\frac{a^({2)}}{2}$ = 4          &
	                                   &
	                                   &
          -0.158                           &
	                                   &
	                                   &
          -7.569                           \\
          \hline
          \hline
       \end{tabular}
    \end{center}
    \label{table:results-pol}
 \end{table}

 From the knowledge of the change of the macroscopic electric field 
 across the interface we can determine the difference in the zero-field
 polarization of the two materials. Assuming that in our toy model the 
 dielectric constant of the two materials are the same and equal to 1, then

 \begin{equation}
    - 4 \pi \left( \mathcal{P}^{(2)} - \mathcal{P}^{(1)} \right)
    = \epsilon \left( \mathcal{E}^{(2)} - \mathcal{E}^{(1)} \right).
 \end{equation}

 \noindent Since $\mathcal{E}^{(2)} - \mathcal{E}^{(1)} = 4 \pi \sigma_{pol}$,

 \begin{equation}
    - \left( \mathcal{P}^{(2)} - \mathcal{P}^{(1)} \right) = \sigma_{pol}.
 \end{equation}

 \noindent But our material 2 is unpolarized, so $\mathcal{P}^{(2)} = 0$,
 and we arrive to the conclusion that 

 \begin{equation}
    \mathcal{P}^{(1)} = \sigma_{pol}.
 \end{equation}

 This conclusion can be checked analytically in our particular toy model,
 where the charge density is the juxtaposition of ``atomic-like '' charge
 densities. Integrating the first-moment of the microscopic charge
 density for the {\it slab} of material 1 used to build the superlattice,
 defined in Eq. \ref{eq:combgauss}), we arrive to the conclusion that
 $\mathcal{P}^{(1)} = -\frac{\delta^{(1)}}{a^{(1)}_{bulk}}$, 
 whose numerical value in our numerical example equals the polarization
 charge shown in Table \ref{table:results-pol}.
 A generalization for the dielectrically mismatched interface can be found in 
 Sec. III. F of Ref. \onlinecite{Vanderbilt-93}.

\begin{figure}[htbp]
    \begin{center}
       \includegraphics[width=10cm]{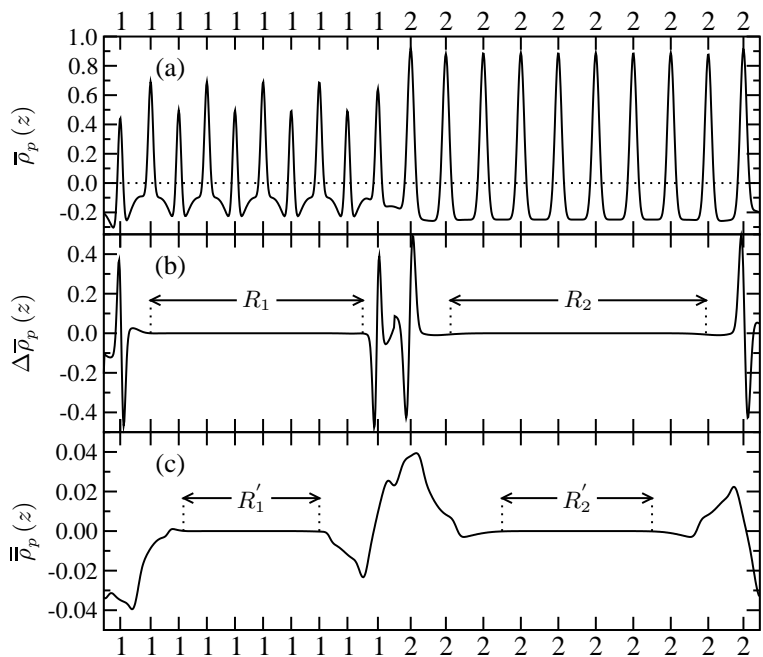}
       \caption{ (a) Microscopic charge density 
                 $\overline{\rho}_{p} \left( z \right)$ within a unit supercell
                 for a polarized system in which the unit cell of 
                 material 1 has two atoms of opposite charge in the unit cell.
                 The widths of the layers of both material and the rest of
                 the parameters of the toy model are as in the caption of Fig.
                 \ref{fig:nanotoypolar}, with the exception of 
                 $A^{e,1} = 1.5$ (net charge, -0.5 per site) for the odd atoms
                 and $A^{e,1} = 0.5$ (net charge, +0.5 per site) for the even
                 atoms of material 1, 
                 and $\delta^{(1)}_{i} = 0.0$, 
                 where $i$ runs over all the ``atoms''
                 of material 1. 
                 (b) Interface-induced deformation of the charge density 
                 $\Delta \overline{\rho}_{p} \left( z \right)$ 
                 defined as in Eq. (\ref{eq:deltarhop}).
                 As before, the interface plane $z_{int}$ is positioned at
                 the middle of the
                 separation between the last atomic plane on the
                 left and the first atomic plane on the right.
                 (c) Nanosmoothed charge density 
                 $\overline{\overline{\rho}}_{p}(z)$ of the microscopic charge 
                 density. The nanosmoothing function is defined as in 
                 Eq. (\ref{eq:filterconvolution}) with 
                 $l_{1} = a^{(1)} = 6$, and 
                 $l_{2} = a^{(2)} = 8$,  single interplanar 
                 distances along $z$ for each material.
                 Atomic units are used.
         }
       \label{fig:nanotoypolarbasis}
    \end{center}
 \end{figure}

 A second toy model we shall analyze for the polarized case is shown in 
 Fig. \ref{fig:nanotoypolarbasis}. It is made of a one dimensional chain
 in which the unit cell of material 1 has two atoms per unit cell.
 Inside the unit cell, one of the atoms is charged negatively and the other
 positively so that the overall charge in the unit cell is neutral.
 Thus, besides the electronic polarization discussed in the previous case,
 polarization can be induced by ionic displacements.
 The parameters of the model are as in Fig. \ref{fig:nanotoypolar},
 with the exception of $A^{e,1} = 1.5$ for the odd ``atoms'' 
 (net charge, -0.5 per site), $A^{e,1} = 0.5$ for the even ``atoms''
 (net charge, +0.5 per site), and $\delta^{(1)}_{i} = 0.0$,
 where $i$ runs over all the ``atoms''
 of material 1 (no displacement of the electronic cloud is considered).

 The main conclusions that can be drawn from Fig. \ref{fig:nanotoypolarbasis}
 are the same as in Fig. \ref{fig:nanotoypolar}.
 As we made for the electronic polarized case, we can infer the value
 of the bulk zero field polarization from the polarization charge at the
 interface, that in this case amounts to +0.25.

 \section{Summary and discussion.}
 \label{sec:summary}

 The problem of how to extract from the immense detail provided by the
 first-principles calculations, with resolution at the atomic scale, 
 reliable values of physical quantities of interest which enter into
 nanoscale electrostatic analysis has been reviewed.

 This problem is particularly challenging in the case of interfaces between
 quite different materials, since all the relevant magnitudes 
 (interface charge and dipole densities, screening lengths, etc.) are
 overwhelmed by the large and rapid oscillations of the microscopic
 charge density.

 The different procedures to filter out the periodic oscillations of the 
 microscopic quantities, which typically follow the underlying atomic
 structure, preserving only those features that change in the vicinity
 of a surface or interface are critically analyzed, and the criteria
 under which they accurately extract the quantities of interest are discussed.

 The approach of defining a reference interface charge density from the
 bulk unperturbed charge densities of each material is spoilt by the fact that
 the interface position is undetermined. The profile of the charge density
 at the interface is discontinuous and displays large fluctuations in the 
 neighborhood of the interface due to the interface-induced atomic
 relaxations. Moreover, both the interface charge and dipole
 densities (and so the potential drop at the interface) 
 are not unique in this approach, since they depend critically on the position
 of the interface.
 
 A clearly superior method consists in nanosmoothing the microscopic
 charge density by taking its convolution with a filter function.
 This procedure eliminates the contributions coming from the bulk
 together with the oscillations due to the interface-induced relaxations 
 while producing a continuous charge distribution.

 In this work, we prove rigorously that
 the interface charge and dipolar densities are independent of the
 nanosmoothing function used provided the following conditions are met:
 $(i)$ it is positive definite, even and normalized within the support
 region of space within which it is defined 
 [Eqs. (\ref{eq:filterfuna})-(\ref{eq:filterfund})];
 $(ii)$ it must be sufficiently smooth itself for the Poisson equation
 to remain invariant after nanosmoothing. By ``sufficiently smooth'' 
 we understand that the second derivative must exist, and both the function
 and the first derivative must vanish at the end points.
 [Eqs. (\ref{eq:conditionf1})-(\ref{eq:conditionf3})];
 $(iii)$ in practice the convolution is done using fast Fourier transforms.
 Therefore, although the nanosmothing function might violate
 some of the previous conditions, the family of the finite Fourier series
 involved in the convolutions is a distribution which converges
 to the smoothing function in the limit, meeting all the requirements
 along the way.
 $(iv)$ if square-wave filter functions (or convolutions of them)
 are chosen as filtering functions, the width of each filter function
 $l_{1}$ and $l_{2}$ must be integral numbers of the lattice constant of
 each material in order to have charge and dipole moment densities 
 (and therefore potential shifts across the interface)
 insensitive to the smoothing function;
 $(v)$ the smallest acceptable value for $l_{1}$ and $l_{2}$  
 should be used in order to avoid the presence of multiple maxima
 and minima in the smoothed charge density;
 $(vi)$ the total width of the filter function $L$ should be chosen
 on the scale of the unit cell length
 or larger, but significantly smaller than the regions $R_{s}$
 of each material where the microscopic charge density is unaffected
 by the presence of the interface;
 $(vii)$ after nanosmoothing, the charge density displays two regions
 $R_{s}^{'}$, within the $R_{s}$, where the smoothed charge density 
 $\overline{\overline{\rho}}$ vanishes. The integration limits
 $z_{1}$ and $z_{2}$ used to compute the interface charge and dipole densities
 must lie in these regions;
 $(viii)$ once the integration limits are chosen, 
 the charge density at the interface computed from the microscopic and
 the nanosmoothed charge density are equal if and only if the microscopic
 charge density is symmetric for $\left| z - z_{1,2} \right| \le L$.
 The interface dipole density computed from the microscopic and
 the nanosmoothed 
 charge density are never equal;
 $(ix)$ nanosmoothing is only valid for computing charge and dipole moment
 densities. Nothing can be said about the shape of the nanosmoothed charge
 density at the interface since it depends critically on the
 filter function;
 $(x)$ by using the smallest acceptable values of $l_{1}$ and $l_{2}$,
 generally one unit cell length, reasonable estimates
 of the density of the charge transferred across the interface and of 
 dipole-layer width can be made.

 The nanosmoothing procedure is a powerful technique that allows us
 to extract the information relevant for computing the change in the
 average potentials and charge densities from one side of the interface 
 to the other, opening the door to the calculations of band offsets,
 \cite{Baldereschi-88,Colombo-91,Peressi-98,Franciosi-96} 
 polarization \cite{Giustino-03} and dielectric permittivity \cite{Giustino-05}
 profiles, 
 effective charges, \cite{Martin-81} and
 force constants \cite{Kunc-82} in semiconductor-semiconductor interfaces,
 and depolarizing electric fields and screening lengths in 
 real ferroelectric capacitors. \cite{Junquera-03.1,Ghosez-06}
 The present work sets the basis for a rigorous selection of the 
 minimum size of the supercell required to obtain accurate
 and smoothing-independent results.

\begin{acknowledgments}
  The authors are indebted to Richard M. Martin for very clarifying 
  discussions.
  JJ acknowledges finantial support of the Spanish Ministery of Science and
  Education through the ``Ram\'on y Cajal'' program, 
  MEC Grants. No. MAT2005-00107 and FIS2006-02261, 
  the NSF MRSEC Grant No. DMR-00-80008, and the
  ARC Discovery Grant DP 0666231.
\end{acknowledgments}
 
\appendix
\section{Transverse averaging of the Poisson equation.}
\label{sec:transversepoisson}

 The microscopic Poisson equation is written as

 \begin{equation}
    \nabla^{2} V \left( \vec{r} \right) = - 4 \pi \rho \left( \vec{r} \right). 
    \label{eq:micropoisson}
 \end{equation}

 \noindent Integrating both sides of Eq. (\ref{eq:micropoisson}) in the 
 $\left( x,y \right)$ plane and dividing by the surface $S$ of the interface
 unit cell, we get

 \begin{align}
    \frac{1}{S} \int \int_{S}  \nabla^{2} V \left( \vec{r} \right) 
    \: dx \:dy & =
    - 4 \pi \frac{1}{S} \int \int_{S}  \rho \left( \vec{r} \right)
    \: dx \:dy 
    \nonumber \\
    & = - 4 \pi \overline {\rho} \left( z \right) .
    \label{eq:planpoisson}
 \end{align}

 \noindent We now write out the integral of the left-hand-side 
 of Eq. (\ref{eq:planpoisson}) explicitly,

 \begin{align}
    \frac{1}{S} \int \int_{S}  \nabla^{2} V \left( \vec{r} \right) 
    \: dx \:dy & =
    \frac{1}{S} \int \int_{S} \left( 
    \frac{ \partial^{2} }{ \partial x^{2} } + 
    \frac{ \partial^{2} }{ \partial y^{2} } + 
    \frac{ \partial^{2} }{ \partial z^{2} }  
    \right)
    V \left( \vec{r} \right) \: dx \:dy 
    \nonumber \\
    & = 
    \frac{1}{S} \int \int_{S} 
    \frac{ \partial^{2} V \left( \vec{r} \right) }
    { \partial x^{2} }  
    \: dx \:dy +
    \frac{1}{S} \int \int_{S} 
    \frac{ \partial^{2} V \left( \vec{r} \right) }
    { \partial y^{2} }  
    \: dx \:dy + 
    \frac{1}{S} \int \int_{S} 
    \frac{ \partial^{2} V \left( \vec{r} \right) }
    { \partial z^{2} }  
    \: dx \:dy .
    \label{eq:decompxyz}
 \end{align}

 \noindent The first two integrals on the right-hand-side 
 can be performed trivially,

 \begin{align}
   \frac{1}{S} \int \int_{S} 
    \frac{ \partial^{2} V \left( \vec{r} \right) }
    { \partial x^{2} }  
    \: dx \:dy & = 
   \frac{1}{S} \int 
    \left. \frac{ \partial V \left( \vec{r} \right) }
    { \partial x }  \right|_{boundary \: x} dy = 0,
 \end{align}

 \noindent where $boundary \: x$ refers to the two intersections of the 
 boundary of the unit cell with the $x$ axis.
 Since the potential and all its derivatives are periodic in the plane,
 the previous integral vanishes. 
 The same holds for the second integral in the right-hand-side
 of Eq. (\ref{eq:decompxyz}). 
 
 Regarding the third integral in Eq. (\ref{eq:decompxyz}),
 we can take the second derivative with respect $z$ out of the 
 integral,

 \begin{align}
    \frac{1}{S} \int \int_{S} 
    \frac{ \partial^{2} V \left( \vec{r} \right) }
    { \partial z^{2} }  
    \: dx \:dy & =
    \frac{ \partial^{2} } { \partial z^{2} }  
    \left( 
    \frac{1}{S} \int \int_{S} V \left( \vec{r} \right) \: dx \:dy 
    \right)
    \nonumber \\
    & = \frac{ \partial^{2} \overline{V} \left( z \right) } { \partial z^{2} } =
        \frac{ d^{2} \overline{V} \left( z \right) } { d z^{2} } .
    \label{eq:derz}
 \end{align}

 Gathering together the results of Eqs. (\ref{eq:planpoisson}) 
 through (\ref{eq:derz}) we can conclude that

 \begin{equation}
  \overline{\nabla^{2} V \left( \vec{r} \right) }  
    = \nabla^{2} \overline{V} \left( z \right) 
    = \frac{ d^{2} \overline{V} \left( z \right) } { d z^{2} } 
    = - 4 \pi \overline{\rho} \left( z \right);
 \end{equation}

 \noindent that is, the transverse average of the Poisson equation
 yields a one-dimensional Poisson equation for the transverse 
 average of the potential. 

\section{Nanosmoothing of the Poisson equation.}
\label{sec:nanosmoothpoisson}

 Applying the nanosmoothing procedure to the electrostatic potential
 $V \left( \vec{r} \right)$ results in

 \begin{equation}
    \overline{\overline{V}} \left( z \right) =
    \int_{z-L}^{z+L} dz^{'} f ( z - z^{'} ) \overline{V} \left( z \right),
 \end{equation}

 \noindent where $\overline{V} \left( z \right)$ is the planar
 average of the of $V \left( \vec{r} \right)$ in planes
 parallel to the interface,

 \begin{equation}
    \overline{V} \left( z \right) = \frac{1}{S} 
    \int_{S} V \left( \vec{r} \right) \: dx \: dy,
 \end{equation}

 \noindent and $f( z - z^{'})$ vanishes for $z^{'} \geq z+L$ 
 and $z^{'} \leq z-L$. 
 Taking the second derivative of the nanosmoothed potential
 and requiring that $\frac{d^{2}f ( z - z^{'})}{dz^{2}}$ exists yields

 \begin{align}
    \nabla^{2} \overline{\overline{V}} \left( z \right)&  = 
    \frac{d^{2}  \overline{\overline{V}} \left( z \right)}{dz^{2}} = 
    \nonumber \\
    & = \int_{z-L}^{z+L} dz^{'} \frac{d^{2}}{dz^{2}} \left[ 
        f ( z - z^{'} ) 
        \overline{V} ( z^{'} ) \right]
    \nonumber \\
    & = \int_{z-L}^{z+L} dz^{'} \: \frac{d^{2} f ( z - z^{'} )}{dz^{2}} \:
        \overline{V} ( z^{'} ) 
    \nonumber \\
    & = \int_{z-L}^{z+L} dz^{'} \: 
        \frac{d^{2} f ( z - z^{'} )} { d {z^{'}}^{2} } \:
        \overline{V} ( z^{'} ) 
    \nonumber \\
    & = \left. \frac{d f ( z - z^{'} )}{d z^{'}} 
    \overline{ V } ( z^{'} ) \right|_{z-L}^{z+L} - 
    \int_{z-L}^{z+L} dz^{'} \: \frac{d f ( z - z^{'} )} { d z^{'} } \:
    \frac{d \overline{V} ( z^{'} ) }{ d z^{'} }.
    \label{eq:nanopoisson-1}
 \end{align}

 \noindent Requiring that 

 \begin{equation}
    \left. \frac{ d f (z)} {d z} \right|_{-L} =
    \left. \frac{ d f (z)} {d z} \right|_{+L} = 
    0,
 \end{equation}

 \noindent makes the first term in the right hand side of 
 Eq. (\ref{eq:nanopoisson-1}) 
 vanish.
 Integrating the second term in the right hand side of that equation
 yields  

 \begin{align}
    & - \int_{z-L}^{z+L} dz^{'} \: \frac{d f ( z - z^{'} )} { d z^{'} } \:
    \frac{d \overline{V} ( z^{'} ) }{ d z^{'} } = 
    \nonumber \\
    & - \left. f ( z - z^{'} )  \frac{d \overline{V} ( z^{'} ) }
    { d z^{'} } \right|_{z-L}^{z+L} +
    \int_{z-L}^{z+L} dz^{'} \: f ( z - z^{'} ) \:
    \frac{d^{2} \overline{V} ( z^{'} ) }{ d {z^{'}}^{2} },
    \label{eq:nanopoisson-2}
 \end{align}

 \noindent so that we arrive at the conclusion that

 \begin{equation}
    \nabla^{2} \overline{\overline{V}} \left( z \right)  = 
    \frac{d^{2}  \overline{\overline{V}} \left( z \right)}{dz^{2}} = 
    \int dz^{'} \: f ( z - z^{'} ) \:
    \frac{d^{2} \overline{V} ( z^{'} ) }{ d {z^{'}}^{2} } =
    \overline{ \overline{ \frac{d^{2} V (z)}{dz^{2}} } }.
    \label{eq:nanopoisson}
 \end{equation}

\section{Electrostatic potential shift and interface dipole density.}
\label{sec:elecdipole}

 The formal solution of the Poisson equation, 
 Eq. (\ref{eq:poissonnanosmoothed}), is

 \begin{align}
    \overline{\overline{V}} \left( z \right) & =
    - 2 \pi \int dz^{'} | z - z^{'} | \:\:
    \overline{\overline{\rho}} ( z^{'} )
    \nonumber \\
    & = - 2 \pi \left[ \int_{-\infty}^{z} dz^{'} ( z - z^{'} ) \:\:
        \overline{\overline{\rho}} ( z^{'} ) +
                       \int_{z}^{+\infty} dz^{'} ( z^{'} - z ) \:\:
        \overline{\overline{\rho}} ( z^{'} ) \right].
 \end{align}

 \noindent The electrostatic potential shift across the interface is,
 therefore,

 \begin{align}
    \Delta \overline{\overline{V}} = &  
    \overline{\overline{V}} \left( z_{2} \right) - 
    \overline{\overline{V}} \left( z_{1} \right) 
    \nonumber \\
    = &  4 \pi \int_{z_{1}}^{z_{2}} dz^{'} z^{'} \:\: 
    \overline{\overline{\rho}} ( z^{'} ) 
    \nonumber \\
    & - 2 \pi \left\{ z_{2} \left[ \int_{-\infty}^{z_{2}} 
    dz^{'} \overline{\overline{\rho}} ( z^{'} ) - 
    \int_{z_{2}}^{\infty} 
    dz^{'} \overline{\overline{\rho}} ( z^{'} )
    \right] 
    - z_{1} \left[ \int_{-\infty}^{z_{1}} 
    dz^{'} \overline{\overline{\rho}} ( z^{'} ) - 
    \int_{z_{1}}^{\infty} 
    dz^{'} \overline{\overline{\rho}} ( z^{'} )
    \right] 
    \right\}.
    \label{eq:deltav}
 \end{align}

 The location of the points at $\pm \infty$ are thus far unspecified.
 They can both be chosen at an image of $z_{1}$ or of $z_{2}$ 
 within the unit cell without loss of generality since all the relevant
 computational procedures require periodicity. In that case,
 all four integrals within the square brackets of Eq. (\ref{eq:deltav}) 
 vanish because of overall electrical neutrality and of the choice 
 of $z_{1}$ and $z_{2}$ at the microscopic unit cell boundaries,
 with Eq. (\ref{eq:relshiftdip}) following. 

\bibliographystyle{apsrev}

\end{document}